\documentclass[12pt,a4paper]{article}
\usepackage[T1]{fontenc}
\usepackage{float}
\usepackage{mathrsfs}
\usepackage{amsmath}
\usepackage{graphicx}
\usepackage{esint}

\makeatletter

\providecommand{\tabularnewline}{\\}

\usepackage{jcappub}

\title{Cosmology of the interacting Cubic Galileon}

\author{Sihem Zaabat,}
\author{Khireddine Nouicer}

\affiliation{Laboratory of Theoretical Physics and Department of Physics,\\
Faculty of Exact and Computer Sciences,\\University Mohamed Seddik Ben Yahia,\\
BP 98, Ouled Aissa, Jijel 18000, Algeria}

\emailAdd{khnouicer@univ-jijel.dz}

\abstract{
We consider the cosmological dynamics of the cubic Galileon
field model interacting with dark matter, where the interaction
between the cosmic dark sectors is  proportional to  Hubble
parameter and  dark energy density. The background field equations are converted
into an autonomous dynamical system of first order equations using suitable dimensionless
variables. The fixed points are determined and their stability 
analyzed by means of the eigenvalue method. The analysis of the stability
of the fixed points along that of the avoidance of ghosts and
Laplacias instabilities implies negative coupling parameter.
We find an attractor fixed point in the de Sitter phase, and construct an approximated tracker solution,  which to a
good approximation,  mimics the background cosmic evolution obtained by solving
numerically the autonomous system of equations. In addition, we perform
the statefinder and $Om$ diagnostics and show that for some appropriate
values of the coupling constant and initial conditions, the trajectories in the statefinder
parameter phase space evolves towards a stable state corresponding
to  $\Lambda\textrm{CDM}$ model. The comparison with the observed
distance modulus and Hubble parameter data sets clearly indicates that the
cosmological dynamics of the interacting cubic model is compatible with the existence of an interaction
between the dark sectors.} 


\makeatother

\begin{document}
\maketitle

\section{Introduction}

In recent years, the great amount with increasing  precision of cosmological
observations test of the standard $\Lambda\textrm{CDM}$ cosmological
model strongly indicate the existence of non-baryonic cold dark matter
(CDM), and that the present universe is in an epoch of accelerated
expansion with redshift $z\lesssim1$ \cite{A. G. Riess et al. 1,A. G. Riess et al. 2,S. Perlmutter et al.,D. N. Spergel. ,D. J. Eisenstein et al.,P. Astier et al.,Planck}.
Among many other puzzling problems in $\Lambda\textrm{CDM}$ model,
the recent accelerated expansion of the universe, known as the late-time
acceleration, constitute one of the greatest challenges of modern
cosmology. Many extensions of general relativity have been explored
to alleviate this problem. Given that matter with positive pressure
generates deccelerated expansion, cosmologists suggested that the
late-time acceleration of the universe is sourced by an exotic energy
component with negative pressure, known as \textit{dark energy} (DE)
\cite{Copeland,E.V. Linder,A. Silvestri and M. Trodden,V. Sahni and A. A. Starobinsky}.
One of the simplest and earliest candidate of DE, frequently invoked,
is the cosmological constant $\Lambda$. Even though this model fits
very well the cosmological observations, it is plagued several drawbacks,
like \textit{fine tuning }and \textit{coincidence} \textit{problem}
\cite{P. J. Steinhardt}. The first one, known as the \textit{cosmological
constant problem}, is related to the large discrepancy between the
theoretical value of $\Lambda$ predicted by quantum field theories
and its observed value \cite{S. Weinberg}. The \textit{coincidence
problem} asks the question, ``\textit{Why are the densities of non
relativistic matter and vacuum of the same order precisely today}
?\textquotedblright , even though they evolve differently. These problems
have led to increased interest in models where general relativity
is modified in a way to alleviate the mentioned problems and produce
the observed late time acceleration without generating new drawbacks.
A plethora of theoretical models have been proposed to solve the mentioned
problems. Two ways have been borrowed, the first one consists in
modeling the right hand side of Einstein's equations with specific
forms of energy-momentum tensor containing the new exotic energy component
with negative pressure, and an equation of state parameter, $w_{\textrm{de}}<-1/3$.
If the equation of state is not just a constant, the fine tuning problem
usually plaguing DE models with constant equation of state, can be avoided.
An important class of DE models having this property are scalar quintessence
field \cite{Y. Fujii,L. H. Ford,C. Wetterich,B. Ratra and J,Y. Fujii and T. Nishioka,R. R. Caldwell},
where the quintessence field couples minimally to gravity and is a
dynamical slowly evolving component with negative pressure, and its
equation of state is no longer constant but slightly evolves with
time. It behaves like a cosmological constant, $\omega\approx-1$,
if the potential term dominates over the kinetic term and thus generates
sufficient negative pressure to drive the late time acceleration.
However, attempts to resolve the coincidence problem are faced with
fine tuning of the model parameters. Another variety of scalar field
DE models has been proposed including k-essence \cite{C. Armendariz-Picon},
and tachyons fields \cite{Sen,A. Sen} and Chaplygin gaz \cite{chap1,chap2}. 

Because of the infeasibility of solving such theoretical problems,
cosmologists have started to look at the second approach, which tends
to modify the left hand side of Einstein's equation. This alternative
approach postulates that general relativity is only accurate on small
scales, and the laws of gravity fail to describe the universe at large
scales, so it has to be modified. One of the best studied models in
this approach is the 5-dimensional Dvali- Gabadadze-Parroti brane-world
model (DGP) \cite{Dvali}, in which matter is confined to the 4-dimensional
brane and only gravity can propagate in the 5-dimensional bulk. The
DGP model can be compatible with local gravity constraints. However,
its self-accelerating solution contains a ghost mode \cite{A. Nicolis,Gorbunov},
in addition to being incompatible with the joint observational constraints
of Supernovae Ia (SN Ia), Baryon Acoustic Oscillations (BAO) and Cosmic
Microwave Background (CMB) anisotropies \cite{Koyama,Sawicki,Fairbairn,Maartens,Alam,Song}.
In order to avoid the appearance of ghosts in DGP model, a solution
to the large scale modification of gravity is provided by an effective
scalar field $\pi$ dubbed \textit{Galileon}, which appears in the
decoupling limit of DGP model \cite{M. A. Luty and M. Porrati and R. Rattazzi,A. Nicolis and R. Rattazzi},
and satisfying the Galilean shift symmetry $\partial_{\mu}\pi\rightarrow\partial_{\mu}\pi+b_{\mu}$
on Minkowskian background. Nicolis et al. \cite{A. Nicolis} showed
that there are only five field Lagrangians $\mathcal{L}^{\left(i\right)}(i\ =\overline{1,5})$
that respect the Galilean symmetry in the Minkowski background. In
Refs. \cite{Deffayet,C. Deffayet} these Lagrangians have been extended
to covariant forms in curved space-time while keeping the equations
of motion up to second order. Consequently, this property allows to
avoid the appearance of extra unphysical degenerated modes associated
with ghosts, while the late-time cosmic acceleration can be realized
by the field kinetic energy \cite{N. Chow and J. Khoury,T. Kobayashi,T. Kobayashi and  H. Tashiro and D. Suzuki,R. Gannoudji and M. Sami,A. Ali and R. Gannoudji and M. Sami,D. F. Mota and M. Sandstad and T. Zlosnik,A. de Felice and S. Mukhoyama and S. Tsujikawa,A. De Felice and S. Tsujikawa,S. Nesseris and A. De Felice,A. De Felice_PRL}.
There exists also another class of theories of large scale modification
of gravity. These are known as $f(R)$ gravity, where $f$ is a function
of Ricci scalar \cite{Capozziello,S. Capozziello,S. M. Carroll,S. Nojiri}.
Although such theories are able to describe the recent late-time accelerated
expansion on cosmological scales correctly, they typically give rise
to strong effects on local scales. Other models in the same context
are scalar-tensor theories \cite{Amendola,J. P. Uzan,T. Chiba,N. Bartolo,F. Perrotta,A. Riazuelo},
and Gauss-Bonnet gravity \cite{Nojiri,S. Nojiri and S}. An attractive
feature of these models is that the cosmic late-time acceleration
can be realized without recourse to an exotic energy component. 

Finally, a third way to solve the drawbacks of the standard $\Lambda\textrm{CDM model}$
has emerged some times ago and several authors have invoked the possibility,
that DE might directly interact with DM by exchanging energy during
their cosmic evolution. Originally, this approach was used to explain
the smallness of the cosmological constant \cite{wetterichA,wetterichB},
and has become a fruitful framework to solve the \textit{coincidence
problem }\cite{coincid1,coincid2,coincid3,coincid4,coincid5,coinci6,coinci7,coincid8,coincid9,coincid10,coincid11,coincid12,coincid13,coincid14,coincid15}\textit{.}
An other motivation of introducing an interaction between the cosmic
dark sectors is to improve the prediction of the standard cosmological
model concerning the structures formation and their evolution. Inspired
by string theories and scalar-tensor theories, several authors studied
interaction between quintessence field and dark matter \cite{amendola1999,amendola2000}.
Despite the lack of a fundamental theory for the interaction between
dark sectors, a plethora of phenomenological forms of the interaction
term have been proposed and constrained by observations (see e.g.
\cite{L. Santos} and references therein). Recently in \cite{E. Di Valentino},
it has been find that DE-DM interaction alleviate the current tension
between the value $H_{0}$ constrained from the Planck Cosmic Microwave
Background measurement \cite{N. Aghanim} and the recent estimate
of Riess et al.\cite{A. G. Riess2}.

In the present work, we study the cosmological effects of DE-DM interaction
in the framework of the Galileon model. This paper is organized as
follows. In section \ref{sec:Cosmological-Galileon} we present a
short review of the cubic Galileon field model and derive the field
equations on general space-time. In section \ref{sec:Background-dynamics}
we introduce the DE-DM interaction term into the cubic Galileon model
and obtain the background field equations on spatially flat Friedmann-Lemaitre-Robertson-Walker
(FLRW) space-time. Assuming the existence of de Sitter (dS) stable
epoch and exploiting its cosmological properties, we reduce the dimension
of the model parameter space. In section \ref{sec:Generalized-dynamical-system}
using suitable dimensionless dynamical variables the field equations
are converted into an autonomous system of first order differential
equations. The later are solved and the fixed points determined and
their stability analyzed. In \ref{sec:Ghosts-and-laplacian}, the
conditions for the avoidance of ghosts and Laplacian instabilities
are established. In section \ref{sec:Analysis-of-the} we analyze
in detail the cosmological evolution of the background parameters
in the eras governed by the fixed points, along with the no-ghost
and absence of Laplacian instabilities conditions, and derive appropriate
constraints on DE-DM coupling constant. We also show the existence
of a stable dS epoch and construct a approximate tracking solution.
In section \ref{sec:Numerical-Analysis} the detailled numerical integration
of the model is performed for the exact solution and the approximated
tracking solutions. The cosmological behavior is also investigated
using the statefinder and $Om$ diagnostics. Finally, conclusions
are drawn in in Section \ref{sec:Conclusion}.

\section{Cubic Galileon model \label{sec:Cosmological-Galileon}}

Let us start with the action of the cubic Galileon field model as
given in \cite{Deffayet,C. Deffayet}

\begin{equation}
S=\int d^{4}x\sqrt{-g}\left(\frac{M_{\textrm{Pl}}^{2}R}{2}+\frac{1}{2}\sum_{i=1}^{3}c_{i}\mathcal{L}^{\left(i\right)}\left[g_{\mu\nu},\pi\right]+\mathcal{L}_{M}\left[g_{\mu\nu},\left\{ \phi_{a}\right\} \right]\right)\label{Action}
\end{equation}
where $M_{\textrm{Pl}}^{2}$ is the reduced Planck mass related to
Newton's constant $M_{\textrm{Pl}}^{2}=1/8\pi G$, $g$ the determinant
of the metric tensor $g_{\mu\nu}$, $R$ the Ricci scalar, $c_{i}$
are constants parameters, $\mathcal{L}^{\left(i\right)}$ are the
Lagrangians for the Galileon field $\pi$ assumed to be minimally
coupled to the metric, and $\mathcal{L}_{M}$ is the sum of radiation,
baryonic and dark matter contributions with a set of fields $\left\{ \phi_{a}\right\} $.
In this model the Galileon is considered as the source of dark energy.
The Lagrangians of interest are given by 
\begin{equation}
\mathcal{L}^{\left(1\right)}=M^{3}\pi,\quad\mathcal{L}^{\left(2\right)}=\left(\nabla\pi\right)^{2}=\pi_{;\mu}\pi^{;\mu},\quad\mathcal{L}^{\left(3\right)}=\frac{1}{M^{3}}\left(\nabla\pi\right)^{2}\square\pi.\label{L1}
\end{equation}
As we can see, $\mathcal{L}^{\left(1\right)}$ is a linear potential
for the Galileon , $\mathcal{L}^{\left(2\right)}$ the standard quadratic
kinetic Lagrangian, and $\mathcal{L}^{\left(3\right)}$ the cubic
Lagrangian which arises in the decoupling limit of the DGP model \cite{M. A. Luty and M. Porrati and R. Rattazzi,A. Nicolis and R. Rattazzi}.
In addition, $M$ is a constant with dimension of mass, such that
$M^{3}=$$M_{\textrm{Pl}}H_{\textrm{dS}}^{2}$ and $H_{\textrm{dS}}$
being the Hubble expansion rate in the de Sitter (dS) epoch. If we
set $c_{3}=0$ we have the standard quintessence Lagrangian.

The Einstein equations of motion derived by varying the action ($\ref{Action}$)
with respect to $g_{\mu\nu}$ read 
\begin{equation}
M_{\textrm{Pl}}^{2}G_{\alpha\beta}=T_{\alpha\beta},\label{ten2}
\end{equation}
where $G_{\alpha\beta}$ denotes the Einstein symmetric tensor, $T_{\alpha\beta}=\sum_{i=1}^{3}c_{i}T_{\alpha\beta}^{\left(i\right)}+\sum_{a}T_{\alpha\beta}^{\left(a\right)}$
is the total energy-momentum tensor where $T_{\alpha\beta}^{\left(i\right)}$
is the energy-momentum tensor of the Galileon field and $T_{\alpha\beta}^{\left(a\right)}$
the energy-momentum tensor of radiation, baryonic matter and dark
matter $\left(a=\textrm{r,\:b\:,dm}\right)$. Starting from the relation
\begin{equation}
T_{\alpha\beta}^{\left(i\right)}=-\frac{2}{\sqrt{-g}}\frac{\delta S^{\left(i\right)}}{\delta g^{\alpha\beta}},\quad S^{\left(i\right)}=\int d^{4}x\sqrt{-g}\mathcal{L}^{\left(i\right)}\text{ }.
\end{equation}
we easily show that 
\begin{equation}
T_{\alpha\beta}^{\left(1\right)}=M^{3}\pi g_{\alpha\beta}\text{ },\label{eq:t-1}
\end{equation}
\begin{equation}
T_{\alpha\beta}^{\left(2\right)}=-\pi_{;\alpha}\pi_{;\beta}+\frac{1}{2}g_{\alpha\beta}\left(\nabla\pi\right)^{2}\text{ },\label{eq:t-2}
\end{equation}
\begin{equation}
T_{\alpha\beta}^{\left(3\right)}=-\frac{1}{M^{3}}\left(\pi_{;\alpha}\pi_{:\beta}\square\pi+g_{\alpha\beta}\pi_{;\mu}\pi^{;\mu\lambda}\pi_{;\lambda}-\pi^{;\mu}\left[\pi_{;\alpha}\pi_{;\beta\mu}+\pi_{;\beta}\pi_{;\alpha\mu}\right]\right).\label{eq:t-3}
\end{equation}
For matter components, we assume the usual energy-momentum tensor
describing a perfect fluid 
\begin{equation}
T_{\alpha\beta}^{\left(i\right)}=\left(\rho_{i}+p_{i}\right)u_{\alpha}u_{\beta}+p_{i}g_{\alpha\beta}\label{eq:Perfect_Fluid}
\end{equation}
where $\rho_{i}$ is the energy density, $p_{i}$ the pressure, and
$u_{\alpha}$ the velocity of the fluid. We define also an Equation
of State parameter (EoS) for each component in the form $p_{i}=w_{i}\rho_{i}.$
For radiation and pressurless mater (baryonic and cold dark matters)
we have $w_{r}=1/3$ and $w_{b}=w_{dm}=0,$ respectively. In order
to preserve the local energy-momentum conservation law, we have the
conservation law 
\begin{equation}
\nabla_{\alpha}T^{\alpha\beta}=0.\label{eq:Cons_tot}
\end{equation}
Finally, varying the action with respect to the Galileon field we
obtain the equation of motion 
\begin{equation}
\frac{c_{1}M^{3}}{2}-c_{2}\square\pi-\frac{c_{3}}{M^{3}}\left[\left(\square\pi\right)^{2}-R_{\mu\nu}\pi^{;\mu}\pi^{;\nu}-\pi_{;\mu\nu}^{2}\right]=0.\label{eq:eq_galileon}
\end{equation}

\section{Background dynamics\label{sec:Background-dynamics}}

In order to study the cosmological consequences of the interacting
cubic Galileon, we assume the geometry of spatially flat expanding
universe described by the FLRW metric 
\begin{equation}
ds^{2}=-n^{2}\left(t\right)dt^{2}+a^{2}(t)\delta_{ij}dx^{\text{i}}dx^{j}
\end{equation}
where $a\left(t\right)$ is the scale factor. Using $n^{2}\left(t\right)=1,$
the Friedmann equations on the FLRW background are obtained from the
$(0,0)$ and $(i,j)$ components of Einstein equations ($\ref{ten2})$
\begin{equation}
3M_{\textrm{Pl}}^{2}H^{2}=-\frac{1}{2}c_{1}M^{3}\pi-\frac{1}{2}c_{2}\dot{\pi}^{2}+\frac{3}{M^{3}}c_{3}H\dot{\pi}^{3}+\rho_{\textrm{dm}}+\rho_{\textrm{b}}+\rho_{\textrm{r}}\text{ },\label{f1}
\end{equation}
\begin{equation}
M_{\textrm{Pl}}^{2}(3H^{2}+2\dot{H})=-\frac{1}{2}c_{1}M^{3}\pi+\frac{1}{2}c_{2}\dot{\pi}^{2}+\frac{1}{M^{3}}c_{3}\dot{\pi}^{2}\ddot{\pi}-\frac{1}{3}\rho_{\textrm{r}}\text{ ,}\label{f2}
\end{equation}
where $H=\frac{\dot{a}}{a}$ is the Hubble expansion rate, and a dot
denote derivative with respect to time. From Friedmann equations (\ref{f1})
and (\ref{f2}), we identify the effective dark energy density and
pressure of the Galileon field 
\begin{equation}
\rho_{\pi}=-\frac{1}{2}c_{2}\dot{\pi}^{2}+\frac{3}{M^{3}}c_{3}H\dot{\pi}^{3},\label{rho}
\end{equation}
\begin{equation}
P_{\pi}=-\frac{1}{2}c_{2}\dot{\pi}^{2}-\frac{1}{M^{3}}c_{3}\dot{\pi}^{2}\ddot{\pi}.
\end{equation}
In the following we are interested by a late-time cosmic exapnsion
acceleration realized by the field kinetic energy, such that we set
$c_{1}=0$ in the rest of this paper. The Galileon field equation
on the FLRW background is then given by 
\begin{equation}
c_{2}M_{\textrm{Pl}}H_{\textrm{d}S}\left(\dot{\pi}H\left(\alpha-6\right)-2\ddot{\pi}\right)+c_{3}\left(6\dot{\pi}^{2}H^{2}\left(\alpha-3\right)-6\dot{\pi}^{2}\dot{H}-12\dot{\pi}\ddot{\pi}H\right)=0\label{eq:field_equation}
\end{equation}
In order to study the cubic Galileon model in its generality and make
room to the existence of an energy transfer between DE and DM fields
as supported by the observation of galaxy clusters \cite{O. Bertolami,M. Baldi},
we assume that the dark sector components do not evolve separately
but interact with each other. The general way to describe this interaction
is to introduce an energy-momentum exchange term into the conservation
equations as follows 
\begin{equation}
\nabla_{\alpha}T_{\textrm{de}}^{\alpha\beta}=-\nabla_{\alpha}T_{\textrm{dm}}^{\alpha\beta}=Q^{\beta}.\label{eq:interaction}
\end{equation}
where $Q^{\beta}$ is given covariantly by \cite{kodama} 
\begin{equation}
Q_{\textrm{de/dm}}^{\beta}=Qu^{\beta},
\end{equation}
and $u^{\beta}$ is the DE/DM 4-velocity. The function $Q,$ known
as the interaction function between DE and DM, is generally function
of DE and DM densities, Hubble parameter and its derivatives. Assuming
that there is only energy transfer between DE and DM we have $Q_{\textrm{de}}=Q_{\textrm{dm}}=-Q.$
We mention that negative $Q$ indicates that DE decays to DM, whereas
DM decays to DE for positive $Q$. In the FLRW space-time, Eq.(\ref{eq:interaction})
gives 
\begin{equation}
\dot{\rho}_{de}+3H\left(\rho_{de}+P_{de}\right)=Q,\label{pi}
\end{equation}
\begin{equation}
\dot{\rho}_{\textrm{dm}}+3H\rho_{\textrm{dm}}=-Q.\label{dm}
\end{equation}
Once a form of $Q$ is given, the background dynamics is fully determined
by the modified energy conservation equations ($\ref{pi}$) and ($\ref{dm}$)
and the Friedmann equations (\ref{f1}) and (\ref{f2}). Additionally,
the conservation laws for radiation and baryonic components on the
FLRW background read 
\begin{equation}
\dot{\rho}_{\textrm{r}}+4H\rho_{\textrm{r}}=0.\label{r}
\end{equation}
\begin{equation}
\dot{\rho}_{\textrm{b}}+H\rho_{\textrm{b}}=0.\label{b}
\end{equation}

At this stage let us take benefit from the existence of a dS background
characterized by $H=H_{\textrm{dS}}\equiv cst$, $\dot{\pi}=\dot{\pi}_{\textrm{dS}}\equiv cst$,
and fix the free parameters $c_{2},\:c_{3}$ in the Galileon Lagrangian
in terms of the interaction function. Writing the Eqs. ($\ref{f1}),\:(\ref{f2}),\:(\ref{eq:field_equation})$
and ($\ref{eq:interaction})$ in the dS era, and solving the resulting
equations we obtain 
\begin{equation}
x_{\textrm{dS}}^{2}c_{2}=6,\quad x_{\textrm{dS}}^{3}c_{3}=\frac{18H_{\mathscr{\textrm{dS}}}^{3}M_{\textrm{Pl}}^{3}+Q_{\textrm{dS}}}{9H_{\mathscr{\textrm{dS}}}^{3}M_{\textrm{Pl}}^{3}}\label{c2}
\end{equation}

\begin{equation}
\Omega_{\textrm{b},\textrm{dS}}=0,\quad\Omega_{\textrm{dm,dS}}=-\frac{Q_{\textrm{dS}}}{9H_{\textrm{dS}}^{3}M_{\textrm{Pl}}^{2}}.\label{eq:Om_dS}
\end{equation}
where we have set $x_{\textrm{dS}}=\frac{\dot{\pi}_{\textrm{dS}}}{H_{\textrm{dS}}M_{\textrm{Pl}}}$
and normalized $M$ to $M^{3}=M_{\textrm{Pl}}H_{\textrm{dS}}^{2}$
(This give $M\approx10^{-40}M_{\textrm{pl}}$ for $H_{\textrm{dS}}\approx10^{-60}M_{\textrm{Pl}})$
\cite{A. De Felice and S. Tsujikawa}. This clearly shows that $Q_{\textrm{dS}}$
must be negative, signaling an energy transfer from DE to DM in the
dS epoch. It is of interest to remark that the only relevant free
parameters are the coupling parameters in the interaction function
and that DM density in the dS era is not zero but depends on the the
coupling parameters.

\section{Generalized dynamical system\label{sec:Generalized-dynamical-system}}

In this section, we study the cosmological dynamic of the interacting
Galileon through the use of the autonomous dynamical systems and phase
space trajectories analysis. The evolution equations ($\ref{f1}$)
and ($\ref{f2}$ ) are first transformed to first order differential
equations by the introduction of new dimensionless dynamical variables.
To achieve this goal, we follow the analysis performed in the non-interacting
case in \cite{S. Nesseris and A. De Felice,A. De Felice and S. Tsujikawa},
and introduce the dimensionless variables $r_{1}$ and $r_{2}$ 
\begin{equation}
r_{1}=\frac{\dot{\pi}_{\textrm{dS}}H_{\textrm{dS}}}{\dot{\pi}H},\quad r_{2}=\frac{1}{r_{1}}\left(\frac{\dot{\pi}}{\dot{\pi}_{\textrm{dS}}}\right)^{4}.\label{variables}
\end{equation}
From these definitions we easily obtain 
\begin{equation}
\frac{H}{H_{\textrm{dS}}}=\frac{1}{r_{1}\left(r_{1}r_{2}\right)^{1/4}},\quad\frac{\dot{\pi}}{\dot{\pi}_{ds}}=\left(r_{1}r_{2}\right)^{1/4}.\label{eq:Hub_pi}
\end{equation}
In the dS phase, we have $r_{1}=1$ and $r_{2}=1.$ From Eqs. (\ref{r}),
(\ref{b}) and (\ref{variables}) we easily obtain the following differential
equations 
\begin{flalign}
r'_{1}= & -r_{1}\left(\frac{\ddot{\pi}}{\dot{\pi}H}+\frac{H'}{H}\right),\label{eq:ode1}\\
r'_{2}= & r_{2}\left(\frac{\ddot{5\pi}}{\dot{\pi}H}+\frac{H'}{H}\right),\label{eq:ode2}\\
\Omega'_{\textrm{r}}= & -2\Omega_{r}\left(2+\frac{H'}{H}\right),\\
\Omega'_{\textrm{b}}= & -2\Omega_{b}\left(\frac{3}{2}+\frac{H'}{H}\right),\label{eq:ode4}
\end{flalign}
where the prime indicates derivation with respect to $N=\ln a$, and
the dimensionless energy density parameters are defined by 
\begin{equation}
\Omega_{a}=\frac{\rho_{\textrm{a}}}{3M_{\textrm{Pl}}^{2}H^{2}},\quad\textrm{a}=\{\textrm{r,b,dm,de}\}.
\end{equation}
The expression of dark energy density parameter $\Omega_{\textrm{de}}$
is now given in terms of $r_{1}$ and $r_{2}$ as 
\begin{flalign}
\Omega_{\textrm{de}} & =-\frac{r_{1}^{2}r_{2}\left[\left(\alpha-3\right)r_{1}+6-\alpha\right]}{\alpha-3}.\label{Omega_Dark}
\end{flalign}
Combining (\ref{eq:ode1}) and (\ref{eq:ode2}) we obtain the evolution
equation of Hubble parameter 
\begin{equation}
\frac{H'}{H}=-\frac{5r_{1}'}{4r_{1}}-\frac{r_{2}'}{4r_{2}}.\label{eq:Hubble}
\end{equation}
Actually, it is impossible to derive the exact form of $Q$ from first
principles, and the available forms are only on the basis of phenomenological
considerations. Among the various interaction terms studied in the
literature, we take in the present paper, the interaction function
in the form 
\begin{equation}
Q(\rho_{\textrm{dm}},\pi,\dot{\pi})=H\alpha\rho_{\textrm{de}}\label{QQ}
\end{equation}
where $\alpha$ is the constant coupling and $\alpha H$ is the rate
of transfer of DE density. It is well known that this kind of interaction
leads to stable linear perturbations with negative coupling constant$.$
Then, Eqs.(\ref{c2}) and (\ref{eq:Om_dS}) reduce to

\begin{equation}
x_{\textrm{dS}}^{2}c_{2}=6,\quad x_{\textrm{dS}}^{3}c_{3}=\frac{\alpha-6}{\alpha-3},\quad\Omega_{\textrm{b},\textrm{dS}}=0,\quad\Omega_{\textrm{dm,dS}}=\frac{\alpha}{\alpha-3}.\label{c2-1}
\end{equation}

The standard scenario in the non-interacting case is recovered for
$\alpha=0$, and all the analysis performed above reduce to the one
found in \cite{A. De Felice and S. Tsujikawa}. We now use ($\ref{c2-1}$)
and rewrite ($\ref{f1}$) and ($\ref{eq:field_equation}$) in terms
of the dimensionless variables and solve in $\frac{H'}{H}$ and $\frac{\ddot{\pi}}{\dot{\pi}H}$
to obtain 
\begin{flalign}
\frac{H'}{H}= & \left(\alpha-3\right)\frac{\Bigl[r_{1}^{2}r_{2}\left(r_{1}-1\right)\left(\alpha-6\right)^{2}+2\left(3r_{1}^{3}r_{2}-\Omega_{r}-3\right)\left(\alpha r_{1}-\alpha-3r_{1}+6\right)\Bigr]}{\Bigl[4\left(\alpha-3\right)\left(\alpha r_{1}-\alpha-3r_{1}+6\right)-r_{1}^{2}r_{2}\left(\alpha-6\right)^{2}\Bigr]},
\end{flalign}
\begin{flalign}
\frac{\ddot{\pi}}{\dot{\pi}H}= & \frac{\left(\alpha-3\right)\left(\alpha-6\right)\left[3r_{1}^{3}r_{2}-\Omega_{r}-3+2\left(\alpha-3\right)\left(r_{1}-1\right)\right]}{\Bigl[4\left(\alpha-3\right)\left[\alpha\left(r_{1}-1\right)-3r_{1}+6\right]-r_{1}^{2}r_{2}\left(\alpha-6\right)^{2}\Bigr]}.
\end{flalign}
Substituting in Eqs.(\ref{eq:ode1})-(\ref{eq:ode4}), we obtain the
following autonomous system of equations 
\begin{flalign}
r'_{1}= & \frac{r_{1}\left(\alpha-3\right)}{\Delta_{d}}\Bigl[\left(r_{1}^{3}r_{2}-r_{1}^{2}r_{2}+2\left(r_{1}-1\right)\right)\alpha^{2}\nonumber \\
 & +\left(6r_{1}^{4}r_{2}-3r_{1}^{3}r_{2}+12r_{1}^{2}r_{2}-24r_{1}+\Omega_{r}\left(1-2r_{1}\right)+21\right)\alpha\nonumber \\
 & -18\left(r_{1}^{2}-3r_{1}+2\right)r_{1}^{2}r_{2}+54\left(r_{1}-1\right)+6\Omega_{r}\left(r_{1}-1\right)\Bigr],\label{eq:de1}\\
r_{2}'=- & \frac{r_{2}\left(\alpha-3\right)}{\Delta_{d}}\Bigl[\left(r_{1}^{3}r_{2}-r_{1}^{2}r_{2}+10\left(r_{1}-1\right)\right)\alpha^{2}\nonumber \\
 & +\left(6r_{1}^{4}r_{2}-15r_{1}^{3}r_{2}+12r_{1}^{2}r_{2}-96r_{1}-\Omega_{r}\left(3+2r_{1}\right)+81\right)\alpha\nonumber \\
 & -18\left(r_{1}^{2}+r_{1}+2\right)r_{1}^{2}r_{2}+18\left(11r_{1}-7\right)+6\Omega_{r}\left(r_{1}+2\right)\Bigr],\label{eq:de2}\\
\Omega_{\textrm{r}}'= & \frac{2\Omega_{r}}{\Delta_{d}}\Bigl[\left(r_{1}-1\right)r_{1}^{2}r_{2}\alpha^{3}+\left(6r_{1}^{4}r_{2}-21r_{1}^{3}r_{2}+13r_{1}^{2}r_{2}-2\Omega_{r}\left(r_{1}-1\right)+2\left(r_{1}-1\right)\right)\alpha^{2}\nonumber \\
 & +\left(-36r_{1}^{4}r_{2}+126r_{1}^{3}r_{2}-48r_{1}^{2}r_{2}+6\Omega_{r}\left(r_{1}-3\right)-6\left(r_{1}-3\right)\right)\alpha\nonumber \\
 & +54r_{1}^{4}r_{2}-216r_{1}^{3}r_{2}+36r_{1}^{2}r_{2}-18\Omega_{r}\left(r_{1}-2\right)+18\left(r_{1}-2\right)\Bigr],\label{eq:de3}\\
\Omega_{\textrm{b}}'= & \frac{\Omega_{b}}{\Delta_{d}}\Bigl[2\left(r_{1}-1\right)r_{1}^{2}r_{2}\alpha^{3}+\left(12r_{1}^{4}r_{2}-42r_{1}^{3}r_{2}+27r_{1}^{2}r_{2}-4\Omega_{r}\left(r_{1}-1\right)\right)\alpha^{2}\nonumber \\
 & +\left(-36r_{1}^{4}r_{2}+126r_{1}^{3}r_{2}-48r_{1}^{2}r_{2}+12\Omega_{r}\left(r_{1}-1\right)-6\left(2r_{1}-3\right)\right)\alpha\nonumber \\
 & +54r_{1}^{4}r_{2}-216r_{1}^{3}r_{2}+36r_{1}^{2}r_{2}-18\Omega_{r}\left(r_{1}-2\right)+18\left(r_{1}-2\right)\Bigr]\label{eq:de4}
\end{flalign}
where $\Delta_{d}$ is given by 
\begin{equation}
\Delta_{\textrm{d}}=\left(r_{1}^{2}r_{2}-4r_{1}+4\right)\alpha^{2}-12\left(r_{1}^{2}r_{2}-2r_{1}+3\right)\alpha+36\left(r_{1}^{2}r_{2}-r_{1}+2\right).
\end{equation}
Now we express the dark energy EoS parameter $\omega_{\textrm{de}}=\frac{\rho_{\textrm{de}}}{P_{\textrm{de}}}$
and the total effective EoS parameter $\omega_{\textrm{eff}}=-1-\frac{2}{3}\frac{H'}{H}$
in terms of the dynamical variables

\begin{flalign}
w_{\textrm{de}}= & -\frac{\left(\alpha-3\right)}{3\Delta_{\textrm{de}}}\left[\left(2r_{1}-2\right)\alpha^{3}+\left(12r_{1}^{2}-\Omega_{r}-42r_{1}+27\right)\alpha^{2}\right.\nonumber \\
 & -\left(72r_{1}^{2}-12\Omega_{r}-252r_{1}+108\right)\alpha\nonumber \\
 & \left.+108r_{1}^{2}-36\Omega_{r}-432r_{1}+108\right],\label{omega_de-1}
\end{flalign}
\begin{flalign}
w_{\textrm{eff}}= & \frac{1}{3\Delta_{\textrm{eff}}}\left[2r_{1}^{2}r_{2}\left(r_{1}-1\right)\alpha^{3}+\left[\left(12r_{1}^{2}-42r_{1}+27\right)r_{1}^{2}r_{2}-4\Omega_{r}\left(r_{1}-1\right)\right]\alpha^{2}\right.\nonumber \\
 & -\left[\left(72r_{1}^{2}-252r_{1}+108\right)r_{1}^{2}r_{2}-12\Omega_{r}\left(2r_{1}-3\right)\right]\alpha\nonumber \\
 & +\left.108\left(r_{1}^{2}-4r_{1}+1\right)r_{1}^{2}r_{2}-36\Omega_{r}\left(r_{1}-2\right)\right]\label{eq:omega_eff}
\end{flalign}
where 
\begin{flalign}
\Delta_{\textrm{eff}} & =\left(r_{1}^{2}r_{2}-4r_{1}+4\right)\alpha^{2}-12\left(r_{1}^{2}r_{2}-2r_{1}+3\right)\alpha+36\left(r_{1}^{2}r_{2}-r_{1}+2\right),\\
\Delta_{\textrm{de}} & =\Delta_{\textrm{eff}}\left(\left(r_{1}-1\right)\alpha-3r_{1}+6\right).
\end{flalign}
Let us denote the autonomous system (\ref{eq:de1})-(\ref{eq:de4})
by 
\begin{equation}
\dot{\vec{x}}=f\left(\overrightarrow{x}\right),\quad\overrightarrow{x}=\left(r_{1},r_{2},\Omega_{r},\Omega_{b}\right).
\end{equation}
We proceed now to study the cosmological evolution of the interacting
cubic Galileon model. We first determine the fixed or critical points
of the dynamical system of equations and examine their stability during
the cosmic history. The fixed points are the roots of equations $\dot{x}_{i}=0.$
The stability analysis is performed using first order perturbation
technique around the fixed points and then form the matrix of coefficients
of the perturbed terms. A fixed point is said stable (attractor) if
the eigenvalues of the perturbation matrix are all negative, saddle
if the eigenvalues are of mixed signs and unstable if the eigenvalues
are all positive. We have found 5 fixed points, $A,\:B,\:C,\:D$ and
$E$ and summarized their properties in Table.\ref{Tab1}, where $\mathcal{D}_{C}=\left[-\infty,-\frac{15+5\sqrt{17}}{2}\right[\cup\left]-\frac{15-5\sqrt{17}}{2},+\infty\right]$,
$q=-1-\left(H'/H\right)$ is the deceleration parameter, and

\begin{equation}
C_{\pm}=\frac{3\left(\alpha-3\right)}{2}\left[\frac{3\alpha-12\alpha+48\pm\alpha\sqrt{\left(3\alpha-20\right)\left(\alpha-12\right)}}{\alpha^{2}-24\alpha+72}\right].
\end{equation}

\begin{table}
\centering %
\begin{tabular}{c|c|c|c|c}
\hline 
Fixed Points  & $\left(r_{1},\:r_{2}\:,\Omega_{r}\:,\Omega_{b}\right)$  & Eigenvalues  & Stability & q\tabularnewline
\hline 
A  & $\left(0,0,1,0\right)$  & $(1,1,\frac{5-\alpha}{2},-\frac{9-5\alpha}{2})$  & $\begin{array}{c}
\text{ Unst. }9/5<\alpha<5\text{, }\\
\text{Saddle otherwise}
\end{array}$ & $1$\tabularnewline
\hline 
B  & $\left(0,0,0,\Omega_{b}\right)$  & $(0,-1,\frac{9-2\alpha}{4},-\frac{21-10\alpha}{4})$  & $\begin{array}{c}
\text{ Unst. if }\alpha<5/2\text{, }\\
\text{Saddle otherwise}
\end{array}$ & $1/2$\tabularnewline
\hline 
C  & $\left(\frac{\alpha^{2}-11\alpha+30}{\alpha^{2}-13\alpha+30},0,1,0\right)$  & $(8,-\frac{\alpha^{2}}{10}+\frac{3\alpha}{2}-5,1,1)$  & $\begin{array}{c}
\text{ Unst. if }\alpha\in\mathcal{D}_{C}\\
\text{Saddle otherwise}
\end{array}$ & $1$\tabularnewline
\hline 
D  & $\left(\frac{2\alpha^{2}-21\alpha+54}{2\alpha^{2}-24\alpha+54},\:0,\:0,\:\Omega_{b}\right)$  & $(0,-1,6,-\frac{\alpha^{2}}{9}+\frac{3\alpha}{2}-\frac{9}{2})$  & Saddle  & $1$\tabularnewline
\hline 
E  & $\left(1,1,0,0\right)$  & $(-4,-3,\:C_{+},\:C_{-})$  & $\begin{array}{c}
\text{ Stable if }\alpha<3\text{, }\\
\text{Unst. otherwise}
\end{array}$ & $-1$\tabularnewline
\hline 
\end{tabular}\caption{\label{Tab1} Location, Eigenvalues and Stability of the Critical
Points.}
\end{table}

The two fixed points $A$ and $B$ are radiation and matter domination
eras, respectively, with no contribution from dark energy. These points
are unstable for $\alpha<5$ and $\alpha<9/2$, respectively. They
constitute the so called small regime \cite{A. De Felice and S. Tsujikawa,S. Nesseris and A. De Felice}.
The fixed point C and D are also pure radiation and matter dominated
eras, respectively. They are unstable for $5<\alpha<10$ and $9/2<\alpha<9$,
respectively. The last fixed point E is the dS point. It is stable
if $\alpha<3$ and may plays the role of an attractor of the whole
cosmological evolution, independently of the initial conditions on
$r_{1},\,r_{2},\,\Omega_{r},\,\Omega_{b}$. We have two paths for
a valid cosmological evolution. The first (second) path starts from
the unstable dominated radiation era, A (C), continues toward unstable
dominated matter era, B (D), and ends at the dS point, E. It is useful
to note that, for $\alpha=0,$ all the dynamical analysis performed
here reduce to the one in ref.\cite{S. Nesseris and A. De Felice},
and the fixed points A, B, C, D become saddle points while de dS fixed
point remains stable.

\section{Ghosts and laplacian instabilities\label{sec:Ghosts-and-laplacian}}

In order to discuss the stability of theories described by the Lagrangian
$(\ref{Action}$) in the cosmological context, a full treatment of
the linear perturbation theory on the flat FLRW background has been
presented in \cite{A. De Felice and S. Tsujikawa}. Accordingly, the
conditions for the avoidance of ghosts and Laplacian instabilities
have been determined. For the scalar modes the conditions are given
by 
\begin{equation}
Q_{\textrm{S}}\equiv\frac{\omega_{1}\left(4\omega_{1}\omega_{3}+9\omega_{2}^{2}\right)}{3\omega_{2}^{2}}>0,\label{Qs}
\end{equation}

\begin{equation}
C_{\textrm{S}}^{2}\equiv\frac{3\left(2\omega_{1}^{2}\omega_{2}H-\omega_{2}^{2}\omega_{4}+4\omega_{1}\omega_{2}\dot{\omega}_{1}-2\omega_{1}^{2}\dot{\omega}_{2}\right)-6\omega_{1}^{2}\left[\left(1+\omega_{r}\right)\rho_{r}+\left(1+\omega_{dm}\right)\rho_{dm}+\left(1+\omega_{b}\right)\rho_{b}\right]}{\omega_{1}(4\omega_{1}\omega_{3}+9\omega_{2}^{2})}\geqslant0,\label{Cs}
\end{equation}
while the conditions on tensor modes are given by 
\begin{equation}
Q_{\textrm{T}}\equiv\frac{\omega_{1}}{4}>0,\quad C_{\textrm{T}}^{2}\equiv\frac{\omega_{4}}{\omega_{1}}\geqslant0\label{Qt}
\end{equation}
where $C_{\textrm{S}}^{2}$ and $C_{\textrm{T}}^{2}$ are the propagation
speeds squared of scalar and tensor modes, respectively. In the interacting
cubic Galileon model, the functions $\omega_{1},\:\omega_{2},\:\omega_{3}$
and $\omega_{4}$ read

\begin{equation}
\omega_{1}=M_{\textrm{Pl}}^{2},\quad\omega_{2}=-\frac{2c_{3}X\dot{\pi}}{M_{\textrm{Pl}}H_{\textrm{dS}}^{2}}+2M_{\textrm{Pl}}^{2}H,\label{w1}
\end{equation}
\begin{equation}
\omega_{3}=-3c_{2}X+\frac{36c_{3}XH\dot{\pi}}{M_{\textrm{Pl}}H_{\textrm{dS}}^{2}}-9H^{2}M_{\textrm{Pl}}^{2},\quad\omega_{4}=M_{\textrm{Pl}}^{2},\label{w3}
\end{equation}
where $X=-\frac{1}{2}\partial_{\mu}\pi\partial^{\mu}\pi$, and $c_{2}$
and $c_{3}$ given by (\ref{c2-1}). It is worth to note that the
effect of the interaction between dark energy and dark matter is also
encoded in the Hubble parameter. Expressed in terms of $r_{1}$and
$r_{2}$, the functions $\omega_{2}$ and $\omega_{3}$ read

\begin{equation}
\omega_{2}=-\textrm{\ensuremath{M_{\textrm{Pl}}^{2}}}H_{\textrm{dS}}\left[\frac{\left(\alpha-6\right)r_{1}^{2}r_{2}-2\alpha+6}{\left(\alpha-3\right)r_{1}\left(r_{1}r_{2}\right)^{1/4}}\right],\label{w1-2}
\end{equation}
\begin{equation}
\omega_{3}=-9M_{\textrm{Pl}}^{2}H_{\textrm{dS}}^{2}\left[\frac{\left(\alpha-3\right)\left(r_{1}^{3}r_{2}+1\right)-2\left(\alpha-6\right)r_{1}^{2}r_{2}}{\left(\alpha-3\right)r_{1}^{2}\left(r_{1}r_{2}\right)^{1/2}}\right].\label{w3-1}
\end{equation}
Inserting Eqs. ($\ref{w1-2}),$ and ($\ref{w3-1})$ into Eqs. ($\ref{Qs}$),
($\ref{Cs}$), and ($\ref{Qt}$), the conditions for the avoidance
of ghosts and Laplacian instabilities for scalars and tensor modes
become
\begin{flalign}
Q_{\textrm{S}}= & \frac{3\left[(r_{1}^{2}r_{2}-4r_{1}+4)\alpha^{2}-12(r_{1}^{2}r_{2}-2r_{1}+3)\alpha+36\left(r_{1}^{2}r_{2}-r_{1}+2\right)\right]r_{1}^{2}r_{2}}{\left[(\alpha-6)r_{1}^{2}r_{2}-2\alpha+6\right]^{2}},\label{eq:Qs2}\\
C_{\textrm{S}}^{2}=-\frac{1}{3} & \frac{N_{\textrm{S}}}{\left[(\alpha^{2}-12\alpha+36)r_{1}^{2}r_{2}-(4\alpha^{2}-24\alpha+36)r_{1}+4\alpha^{2}-36\alpha+72\right]^{2}},\label{eq:Cs2}\\
Q_{\textrm{T}}= & \frac{M_{\textrm{Pl}}^{2}}{4},\quad C_{\textrm{T}}^{2}=1.
\end{flalign}
where

\begin{flushleft}
\begin{flalign}
N_{\textrm{S}} & =\left(\alpha^{4}-24\alpha^{3}+216\alpha^{2}-864\alpha+1296\right)r_{1}^{4}r_{2}^{2}\nonumber \\
 & +\left(20\alpha^{4}-360\alpha^{3}+2340\alpha^{2}-6480\alpha+6480\right)r_{1}^{5}r_{2}\nonumber \\
 & +\left(-4\alpha^{4}+84\alpha^{3}-648\alpha^{2}+2160\alpha-2592\right)r_{1}^{2}r_{2}\nonumber \\
 & +\left(-48\alpha^{4}+576\alpha^{3}-2592\alpha^{2}+5184\alpha-3888\right)r_{1}^{2}\nonumber \\
 & +\left(8\alpha^{5}-88\alpha^{4}+168\alpha^{3}+1080\alpha^{2}-4752\alpha+5184\right)r_{1}\nonumber \\
 & +\left(-4\alpha^{4}+72\alpha^{3}-468\alpha^{2}+1296\alpha-1296\right)\Omega_{r}\nonumber \\
 & -8\alpha^{5}+124\alpha^{4}-576\alpha^{3}+252\alpha^{2}+3888\alpha-6480
\end{flalign}
It is useful to note that the conditions on the stability of the tensor
modes do not bring additional information on the parameter space. 
\par\end{flushleft}

\section{Analysis of the fixed points\label{sec:Analysis-of-the}}

Let us now proceed to a detailled analysis of the dynamical evolution
in the eras implied by the fixed points, and extract constraints on
the DE-DM coupling from their stability and conditions for the avoidance
of ghosts and Laplacian instabilities.
\begin{itemize}
\item \textbf{\textit{Small regime}} 
\end{itemize}
This regime contains the fixed points A and B where $r_{1}\ll1,\:r_{2}\ll1$.
In this case the autonomous system of equations become 
\begin{flalign}
r_{1}'= & -\frac{r_{1}}{4}\left(2\alpha-\Omega_{r}-9\right),\label{eq:small-r1}\\
r_{2}'= & \frac{r_{2}}{4}\left(10\alpha+3\Omega_{r}-21\right),\label{eq:small_r2}\\
\Omega_{\textrm{r}}'= & \Omega_{\textrm{r}}\left(\Omega_{\textrm{r}}-1\right),\\
\Omega_{\textrm{b}}'= & \Omega_{\textrm{b}}\Omega_{\textrm{r}},
\end{flalign}
which integrate to 
\begin{flalign}
\Omega_{\textrm{r}}= & \frac{1}{1+d_{4}e^{N}},\quad\Omega_{\textrm{b}}=\frac{d_{1}e^{N}}{1+d_{4}e^{N}}\label{eq:small1}\\
r_{1}= & \frac{d_{3}e^{-\frac{N}{2}\left(\alpha-5\right)}}{\left(1+d_{4}e^{N}\right)^{1/4}},\quad r_{2}=\frac{d_{2}e^{\frac{N}{2}\left(5\alpha-9\right)}}{\left(1+d_{4}e^{N}\right)^{3/4}},\label{eq:small2}
\end{flalign}
where $d_{1},\:d_{2},\:d_{3}$ and $d_{4}$ are constants of integration
depending on the cosmological eras considered. The two fixed points
in the small regime, A (unstable for $\alpha<3$) and B (unstable
for $\alpha<5/2$), are pure radiation dominated and pure matter dominated
solutions, respectively. In the vicinity of the point A we can set
$d_{4}\approx0$ and then get from (\ref{eq:small2}) 
\begin{flalign}
r_{1}\approx & a^{\frac{5-\alpha}{2}},\quad r_{2}\approx a^{\frac{5\alpha-9}{2}},\quad H\approx a^{-2},\label{eq:small2-2}
\end{flalign}
while for B we take $d_{4}$ very large but with $d_{1}/d_{4}\sim1$,
such that we get

\begin{flalign}
r_{1}\approx & a^{\frac{9-2\alpha}{4}},\quad r_{2}\approx a^{\frac{10\alpha-21}{4}},\quad H\approx a^{-3/2}\label{eq:small2-2-1}
\end{flalign}
As we can see the exponents of the scale factor in (\ref{eq:small2-2})
and (\ref{eq:small2-2-1}) are exactly the eigenvalues in the fixed
points A and B, respectively, and that $r_{2}$ grows faster than
$r_{1}$ in the interval $9/5<\alpha<5$. We also obtained the usual
evolution laws of the Hubble parameter in the radiation and matter
domination epochs, independently of the DE-DM coupling constant. This
is an expected result, since the interaction term which is proportional
to $\rho_{DE}$, only begins to affect the cosmological evolution
at the onset of DE dominated epoch.

The dark energy and the total effective EoS parameters in the small
regime are now approximated by 
\begin{flalign}
w_{\textrm{de}}\approx & -\frac{1}{12}\left[\frac{2\left(r_{1}^{2}+1\right)\alpha^{2}+\left(2\left(\Omega_{r}+6\right)r_{1}+\Omega_{r}-15\right)\alpha-6\Omega_{r}\left(r_{1}+1\right)+18\left(1-3r_{1}\right)}{\alpha-6}\right],\\
w_{\textrm{eff}}\approx & \frac{\Omega_{r}}{3},
\end{flalign}
and the conditions for the avoidance of ghosts and Laplacian instabilities
become 
\begin{equation}
Q_{\textrm{S}}\approx=\frac{4}{3}\frac{\left(\alpha-6\right)}{\left(\alpha-3\right)}r_{1}^{2}r_{2},\label{eq:small_Qs}
\end{equation}

\begin{equation}
C_{\textrm{S}}^{2}\approx\frac{1}{12}\frac{\left(2\alpha^{2}+\left(\Omega_{r}-7\right)\alpha-6\Omega_{r}-30\right)}{\left(\alpha-6\right)}+\frac{1}{12}\frac{\left(2\alpha^{2}+2\left(\Omega_{r}-2\right)\alpha-6\Omega_{r}-6\right)}{\left(\alpha-6\right)}r_{1}.\label{eq:small_Cs}
\end{equation}
A sign change in $r_{1}$ and $r_{2}$ signals the appearance of ghosts
and Laplacian instabilities for the scalar mode. At exactly the radiation
dominated fixed point A we have 
\begin{flalign}
w_{\textrm{de}}=\frac{1}{6}-\frac{\alpha}{6},\quad w_{\textrm{eff}}=\frac{1}{3},\quad Q_{\textrm{S}}=0,\quad & C_{\textrm{S}}^{2}=\frac{1}{12}\frac{\left(2\alpha^{2}-6\alpha-36\right)}{\left(\alpha-6\right)},\label{eq:small_speed_1}
\end{flalign}
while for the matter dominated fixed point B we have 
\begin{flalign}
w_{\textrm{de}}=\frac{1}{4}-\frac{\alpha}{6},\quad w_{\textrm{eff}}=0,\quad Q_{\textrm{S}}=0,\quad & C_{\textrm{S}}^{2}=\frac{1}{12}\frac{\left(2\alpha^{2}-7\alpha-30\right)}{\left(\alpha-6\right)}.\label{eq:small_speed_2}
\end{flalign}
The condition on Laplacian instabilities for the scalar mode is automatically
satisfied, but the no-ghost condition is only satisfied if $\alpha\in\left[-3,\:6\right[\cup\left]6,\:\infty\right]$
in the radiation dominated small regime, and $\alpha\in\left[-5/2,\:6\right[\cup\left]6,\:\infty\right]$
in the matter dominated small regime, respectively, leading to the
range $\alpha\in\left[-5/2,\:6\right[\cup\left]6,\:\infty\right]$
in the small regime. Additionally, the scalar mode remains sub-luminal
if $\alpha\in\left[-\infty,\:3\right]$ and $\alpha\in\left[-\infty,\:7/2\right]$
in the radiation era and matter era, respectively. Combining these
constraints we must have $\alpha\in\left[-5/2,\:3\right]$ in the
small regime. 

In the non-interacting Galileon model, $\alpha=0$, we have the approximated
solutions $r_{1}\approx a^{5/2},\:r_{2}\approx a^{-9/2},\:H\approx a^{-2},\:w_{\textrm{de}}=1/6$,
and $r_{1}\approx a^{9/4},\:r_{2}\approx a^{-21/4},H\approx a^{-3/2},\:w_{\textrm{de}}=1/4$
for the points A and B, respectively, whereas in \cite{A. De Felice and S. Tsujikawa},
the authors found $r_{1}\approx a^{5/4},\quad r_{2}\approx a^{7/4},\:H\approx a^{-2}$,
and $w_{\textrm{de}}=1/4$ in the radiation era, and $r_{1}\approx a^{9/4},\quad r_{2}\approx a^{3/4},\:H\approx a^{-51/32},$
and $w_{\textrm{de}}=1/8$ for the matter phase. The origin of this
discrepancy lies in numerical errors ($1/8$ in place of $1/4$) in
Eqs. (\ref{eq:small-r1}) and (\ref{eq:small_r2}).
\begin{itemize}
\item \textbf{\textit{Fixed Point C}}
\end{itemize}
This point is unstable in the interval $5<\alpha<10$, and it corresponds
to a pure radiation domination solution. In the vicinity of this point
we have $r_{2}$ very small and then the dark energy density, the
dark energy and the total effective EoS parameters are approximated
by 
\begin{flalign}
\Omega_{\textrm{de}} & =5\left[\frac{\left(\alpha-5\right)^{2}\left(\alpha-6\right)^{3}}{\left(\alpha-3\right)^{3}\left(\alpha-10\right)^{3}}\right]r_{2},\\
w_{\textrm{de}} & =\frac{\alpha}{3}-\frac{7}{3}-\frac{1}{60}\left[\frac{\left(\alpha-6\right)^{3}\left(\alpha-5\right)^{2}\left(\alpha-7\right)}{\left(\alpha-3\right)^{3}\left(\alpha-10\right)}\right]r_{2},\\
w_{\textrm{eff}} & =\frac{1}{3}+\frac{5}{3}\left[\frac{\left(\alpha-6\right)^{3}\left(\alpha-5\right)^{2}\left(\alpha-7\right)}{\left(\alpha-3\right)^{3}\left(\alpha-10\right)^{3}}\right]r_{2}.
\end{flalign}
The conditions for avoidance of ghosts and Laplacian instabilities
read

\begin{equation}
C_{\textrm{S}}^{2}=-\frac{\alpha}{15}+\frac{5}{3}+\frac{1}{300}\left[\frac{\left(\alpha-6\right)^{3}\left(\alpha-5\right)^{3}\left(6\alpha-65\right)}{\left(\alpha-3\right)^{3}\left(\alpha-10\right)}\right]r_{2},
\end{equation}
\begin{equation}
Q_{\textrm{S}}=15\left[\frac{\left(\alpha-5\right)^{2}\left(\alpha-6\right)^{3}}{\left(\alpha-3\right)^{3}\left(\alpha-10\right)^{3}}\right]r_{2}.\label{eq:rad_Qs}
\end{equation}
In this era a sign change in $r_{2}$ signals the appearance of ghosts
and Laplacian instabilities for the scalar mode. At the fixed point
we have $r_{2}=0$, and we are left with 
\begin{equation}
\Omega_{\textrm{de}}=0,\quad w_{\textrm{de}}=-\frac{7}{3}+\frac{\alpha}{3},\quad w_{\textrm{eff}}=\frac{1}{3},\quad C_{\textrm{S}}^{2}=-\frac{\alpha}{15}+\frac{5}{3},\quad Q_{\textrm{S}}=0
\end{equation}
The conditions that $C_{\textrm{S}}^{2}\geq0$ and the scalar mode
remains sub-luminal lead to $\alpha\leq10.$
\begin{itemize}
\item \textbf{\textit{Fixed Point D}}
\end{itemize}
This fixed point corresponds to a pure matter dominated solution. It
is unstable in the interval $9/2<\alpha<9.$ In this era we can expand
the dark energy density, the dark energy and the effective EoS parameters
around $r_{2}\ll1$ to get 
\begin{flalign}
\Omega_{\textrm{de}} & =\frac{9}{8}\left[\frac{\left(2\alpha-9\right)^{2}\left(\alpha-6\right)^{3}}{\left(\alpha-3\right)^{3}\left(\alpha-9\right)^{3}}\right]r_{2},\\
w_{\textrm{de}} & =\frac{\alpha}{3}-2-\frac{1}{216}\left[\frac{\left(2\alpha-9\right)^{2}\left(\alpha-6\right)^{4}}{\left(\alpha-3\right)^{3}\left(\alpha-9\right)}\right]r_{2},\\
w_{\textrm{eff}} & =\frac{3}{8}\left[\frac{\left(2\alpha-9\right)^{2}\left(\alpha-6\right)^{3}}{\left(\alpha-3\right)^{3}\left(\alpha-9\right)^{3}}\right]r_{2}.
\end{flalign}
The conditions for avoidance of ghosts and Laplacian instabilities
now read

\begin{equation}
C_{\textrm{S}}^{2}=-\frac{\alpha}{27}+\frac{4}{3}+\frac{1}{1944}\left[\frac{\left(10\alpha-99\right)\left(2\alpha-9\right)^{2}\left(\alpha-6\right)^{3}}{\left(\alpha-3\right)^{3}\left(\alpha-9\right)}\right]r_{2},
\end{equation}

\begin{equation}
Q_{\textrm{S}}=\frac{27}{8}\left[\frac{\left(2\alpha-9\right)^{2}\left(\alpha-6\right)^{3}}{\left(\alpha-9\right)^{3}\left(\alpha-3\right)^{3}}\right]r_{2}.\label{eq:mat_Qs}
\end{equation}
Again we see that a sign change in $r_{2}$ signals the appearance
of ghosts and Laplacian instabilities for the scalar mode. At exactly
the fixed point we have 
\begin{equation}
\Omega_{\textrm{de}}=0,\quad w_{\textrm{de}}=-2+\frac{\alpha}{3},\quad w_{\textrm{eff}}=0,\quad C_{\textrm{S}}^{2}=\frac{4}{3}-\frac{\alpha}{27},\quad Q_{\textrm{S}}=0.
\end{equation}
The conditions that $C_{\textrm{S}}^{2}\geq0$ and the scalar mode
remains sub-luminal lead to $\alpha\leq9$.
\begin{itemize}
\item \textbf{\textit{dS Fixed Point}}
\end{itemize}
The dS fixed point characterized by $r_{1}=1,\:r_{2}=1$ is stable
for $\alpha<3$. At this point we have

\begin{flalign}
\Omega_{\textrm{de}}=\frac{3}{3-\alpha},\quad w_{\textrm{de}} & =-1+\frac{\alpha}{3},\quad w_{\textrm{\textrm{e}ff}}=-1.
\end{flalign}
Here we observe deviation from the $\Lambda\textrm{CDM }$model. Indeed,
the dS era is still dominated by DE but with a small contribution
from DM. This is the consequence of the form of the interaction term
used in the paper where the DE field decay to DM field. Indeed we
have a small fraction of DM in the dS era given by $\Omega_{\textrm{dm,dS}}=\frac{\alpha}{\alpha-3},$
and that $w_{de}$is not exactly $-1.$

The conditions for the avoidance of ghosts and Laplacian instabilities
for the scalar modes $\left(\ref{eq:Qs2}\right)$, and $\left(\ref{eq:Cs2}\right)$
reduce to 
\begin{equation}
Q_{\textrm{S}}=\frac{3(\alpha^{2}-24\alpha+72)}{\alpha^{2}},\label{eq:dS_Qs}
\end{equation}
\begin{equation}
C_{\textrm{S}}^{2}=-\frac{\alpha\left(5\alpha-12\right)}{3(\alpha^{2}-24\alpha+72)}.
\end{equation}
It is easy to deduce that $Q_{\textrm{S,dS}}>0$ for $\alpha\in\left[-\infty,\:0\right[\cup\left]0,\:12-6\times2^{1/2}\right]\cup\left[12+6\times2^{1/2},\:\infty\right].$
and $C_{\textrm{S,dS}}^{2}\geq0$ for $\alpha\in\left[0,\:12/5\right]\cup\left]12-6\times2^{1/2},\:-12+6\times2^{1/2}\right[.$
The scalar mode remains sub-luminal if $\alpha\in\left[-\infty,\:0\right[\cup\left]0,\:12-6\times2^{1/2}\right[\cup\left]12+6\times2^{1/2},\:\infty\right].$
Then to avoid ghosts and Laplacian instabilities and maintain the
modes sub-luminal, and guarantir the stability of the dS fixed point,
which is realized for $\alpha<3$, the coupling constant must lies
in the interval $\left[0,\:12/5\right].$ 
\begin{itemize}
\item \textbf{\textit{Tracker Solution}}
\end{itemize}
It is assumed that the DE-DM coupling constant $\alpha$ is small,
and we can observe from Table.\ref{Tab1} that the coordinate $r_{1}$
of the fixed points C and D is of order unity. Then following closely
the treatment made in \cite{S. Nesseris and A. De Felice}, we construct
an approximate dynamic of the model by setting $r_{1}=1$ in the dynamical
equations and then solve in terms of $r_{2}$, $\Omega_{r}$ and $\Omega_{b}.$
In this case, the autonomous system of equations (\ref{eq:de1})-(\ref{eq:de4})
become

\begin{flalign}
\frac{r_{2}'}{r_{2}}= & \left(\alpha-3\right)\left[\frac{5\left(\Omega_{r}-3r_{2}+3\right)\alpha-24\left(\Omega_{r}-3r_{2}+3\right)}{\alpha^{2}r_{2}-12\left(\alpha-3\right)\left(r_{2}+1\right)}\right],\label{eq:T1}\\
\frac{\Omega_{\textrm{r}}'}{\Omega_{\textrm{r}}}= & -4\left[\frac{\alpha^{2}r_{2}+3\left(\alpha-3\right)\left(\Omega_{r}-7r_{2}-1\right)}{\alpha^{2}r_{2}-12\left(\alpha-3\right)\left(r_{2}+1\right)}\right],\label{eq:T2}\\
\frac{\Omega_{\textrm{b}}'}{\Omega_{\textrm{b}}}= & -3\left[\frac{\alpha^{2}r_{2}+4\left(\alpha-3\right)\left(\Omega_{r}-6r_{2}\right)}{\alpha^{2}r_{2}-12\left(\alpha-3\right)\left(r_{2}+1\right)}\right].\label{eq:T3}
\end{flalign}
Combining Eqs.(\ref{eq:T1}) and (\ref{eq:T2}) we obtain 
\begin{equation}
\frac{r_{2}'}{r_{2}}-\gamma\frac{\Omega_{r}'}{\Omega_{r}}=4\gamma,\quad\gamma=-\frac{5}{12}\alpha+2.\label{eq:r2/Omr}
\end{equation}
The solution of (\ref{eq:r2/Omr}) is given by 
\begin{equation}
r_{2}\left(N\right)=r_{2}\left(0\right)e^{4\gamma N}\left(\frac{\Omega_{r}\left(N\right)}{\Omega_{r}\left(0\right)}\right)^{\gamma}.\label{eq:r2Sol}
\end{equation}
Also using (\ref{eq:Hubble}) we get 
\begin{equation}
H=H_{0}e^{-\gamma N}\left(\frac{\Omega_{r}\left(N\right)}{\Omega_{r}\left(0\right)}\right)^{-\gamma/4}.
\end{equation}
The DE density, the DE and the effective EoS parameters along the
tracker are now given by
\begin{flalign}
\Omega_{\textrm{de}}= & \frac{3r_{2}}{3-\alpha},\\
w_{\textrm{de}}= & \left(\frac{\alpha-3}{9}\right)\left[\frac{\left(\Omega_{r}+3\right)\alpha^{2}-12\left(\alpha-3\right)\left(\Omega_{r}+6\right)}{\left[r_{2}\alpha^{2}-12\left(\alpha-3\right)\left(r_{2}+1\right)\right]}\right],\\
w_{\textrm{eff}}= & -\left[\frac{r_{2}\alpha^{2}+4\left(\alpha-3\right)\left(\Omega_{r}-6r_{2}\right)}{r_{2}\alpha^{2}-12\left(\alpha-3\right)\left(1+r_{2}\right)}\right].
\end{flalign}
 For $\alpha=0$, we reproduce exactly the relations of \cite{S. Nesseris and A. De Felice}
\begin{flalign}
\textrm{\ensuremath{\Omega_{de}}}= & r_{2},\quad w_{\textrm{de}}=-\frac{\Omega_{r}+6}{3\left(r_{2}+1\right)},\quad w_{\textrm{eff}}=\frac{\left(\Omega_{r}-6r_{2}\right)}{3\left(1+r_{2}\right)}.
\end{flalign}
As we can see the DE density reaches the solution in the dS era for
$r_{2}=1.$ Along the tracker, the conditions for the avoidance of
ghosts and Laplacian instabilities for the scalar modes $\left(\ref{eq:Qs2}\right)$
and $\left(\ref{eq:Cs2}\right)$ reduce to the following relations
\begin{flalign}
Q_{\textrm{S}}= & \frac{3\left(\alpha^{2}r_{2}+12\left(3-\alpha\right)\left(r_{2}+1\right)\right)}{\left(\left(r_{2}-2\right)\alpha-6\left(r_{2}-1\right)\right)^{2}}r_{2},\label{eq:Track_Qs}\\
C_{\textrm{S}}^{2}= & \frac{1}{3\Delta_{\textrm{T}}}\left\lfloor \left(-r_{2}^{2}+4\Omega_{r}-16r_{2}+12\right)\alpha^{4}+\left(24r_{2}^{2}-72\Omega_{r}+276r_{2}-168\right)\alpha^{3}\right.\nonumber \\
 & +\left(-216r_{2}^{2}+468\Omega_{r}-1692r_{2}+1260\right)\alpha^{2}+\left(864r_{2}^{2}-1296\Omega_{r}+4320r_{2}+4320\right)\alpha\nonumber \\
 & \left.-1269r_{2}^{2}+1296\Omega_{r}-3888r_{2}+5184\right\rfloor ,\label{eq:Track_Cs}
\end{flalign}
where 
\begin{equation}
\Delta_{\textrm{T}}=\left[\alpha^{2}r_{2}+12\left(3-\alpha\right)\left(r_{2}+1\right)\right]^{2}.
\end{equation}

Although we can write all the relevant quantities along the tracker
in terms of $\Omega_{\textrm{r}}$, we cannot obtain an algebraic
equation for $\Omega_{r}$ and solve it as done in \cite{S. Nesseris and A. De Felice}.
In our case we have to substitute (\ref{eq:r2Sol}) in (\ref{eq:T2})
and (\ref{eq:T3}) and integrate numerically the resulting system
of equations. Finally, assuming positive DM density in the dS era,
$\Omega_{\textrm{dm,dS}}\geq0,$ we obtain the constraint $\alpha\leq0$.
Based on all the constraints derived above in the various cosmological
epochs, we set in the forecoming numerical analysis $\alpha\leq0,$
even a negative DE-DM coupling constant may induces Laplacian instability
of the scalar perturbation in the dS era.

\section{Numerical Analysis\label{sec:Numerical-Analysis}}

\begin{figure}
\centering \includegraphics[width=7cm,height=5cm]{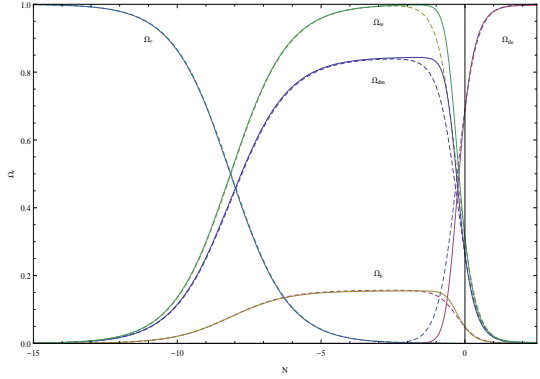}
\caption{\label{Omegas} Background dynamics of the interacting Galileon model.
We plot the evolution of the dimensionless energy densities vs $N$
where the solid lines represent the interacting Galileon model for
$\alpha=-0.0075$ and the initial conditions $r_{1i}=0.0715\times10^{-12},\:r_{2i}=7.17\times10^{-18}.$
The dashed lines represent the energy densities in  $\Lambda\textrm{CDM }$model. }
\end{figure}

\begin{figure}
\centering \includegraphics[width=7cm,height=5cm]{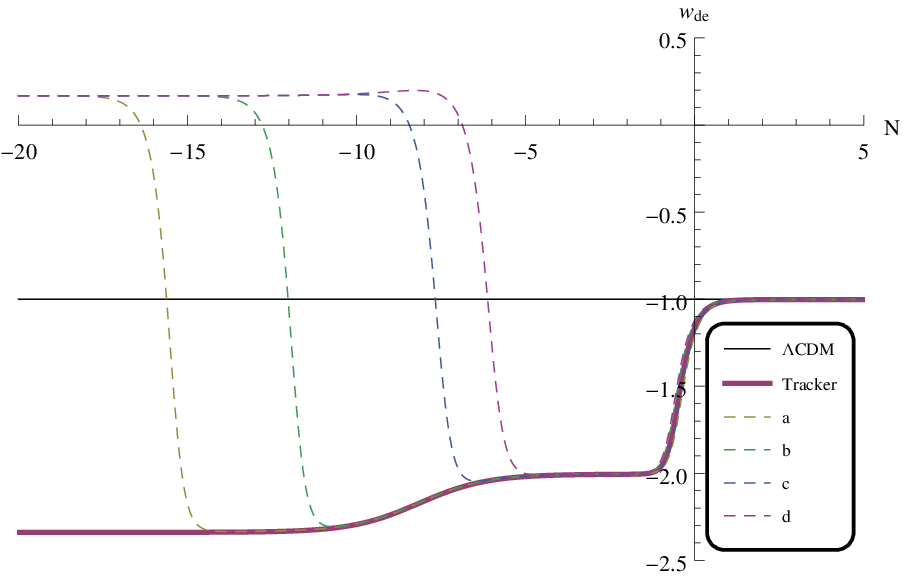}
\hfill{}\includegraphics[width=7cm,height=5cm]{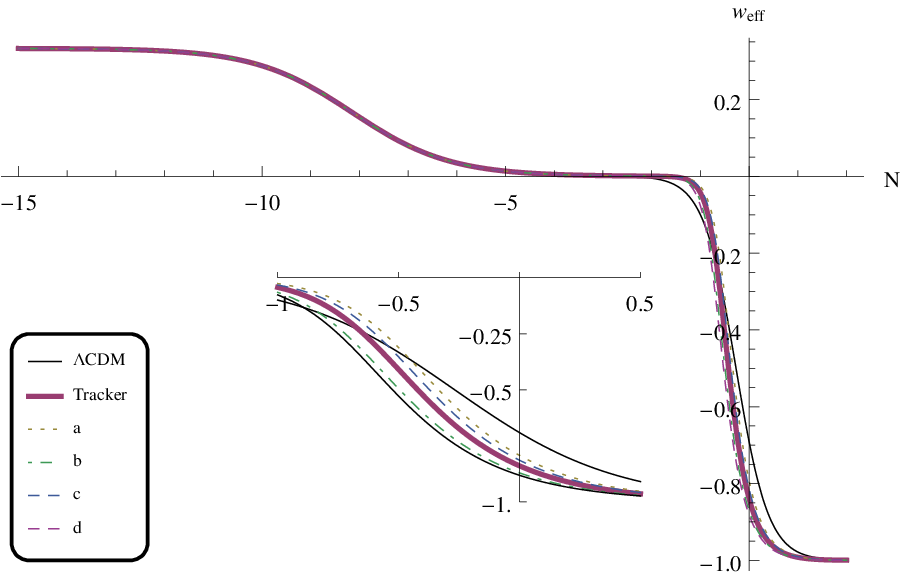} \caption{\label{EoS_parameters} Evolution of $w_{de}$ and $w_{eff}$ vs $N$
for $\alpha=-0.0075$. The different curves represent several initial
conditions at $N=-20$ : a) $r_{1i}=22.6\times10^{-6},\:r_{2i}=0.52\times10^{-38},$
b) $r_{1i}=27.1\times10^{-10},\:r_{2i}=0.442\times10^{-23},$ c) $r_{1i}=0.0715\times10^{-12},\:r_{2i}=7.17\times10^{-18}$,
d) $r_{1i}=22.1\times10^{-16},\:r_{2i}=2.7\times10^{-2}.$ The Thick
curve show the tracker solution with $r_{1}=1$, $r_{2i}=0.71\times10^{-59}$,
and the thin black line is $\Lambda\textrm{CDM }$result.}
\end{figure}

In this section, we integrate numerically the system of Eqs. ($\ref{eq:de1})-(\ref{eq:de4}$)
and confirm the analytic estimation in the epochs implied by the fixed
points and the approximate solution along the tracker discussed in
Sec.\ref{sec:Analysis-of-the}. We need to specify the appropriate
initial conditions to provide the best agreement with the today observations
data as given by the Planck collaboration \cite{Ade}, namely $\Omega_{\textrm{de},0}\simeq0.6910,\:\Omega_{\textrm{m},0}\simeq0.3089,\:H_{0}\simeq67.74\:\textrm{K\ensuremath{m^{-1}s^{-1}}Mpc}$.
In most of our numerical calculation we set the DE-DM coupling constant
to $\alpha=-0.0075$, and set up initially the cosmological evolution
deep in the radiation era in the neighborhood of the fixed point A
at $N=-20$. We also choose the radiation and baryonic initial conditions
as $\Omega_{\textrm{r}i}=0.9999929,\:\Omega_{\textrm{b}i}=1.1013\times10^{-6}$. 
\begin{itemize}
\item \textit{Energy density and EoS parameters} 
\end{itemize}
In Fig.\ref{Omegas} we illustrate the evolution of the density parameters
obtained by integrating numerically Eqs.(\ref{eq:de1}-\ref{eq:de4}).
In comparison with $\textrm{\ensuremath{\Lambda}CDM}$ results shown
by dashed curves, the effect of DE-DM interaction becomes only observable
in the recent past. In Fig.\ref{EoS_parameters} we plot the evolution
of $w_{\textrm{de}}$ and $w_{\textrm{eff}}$ for several initial
conditions, where the thick curve represent the approximated tracker
solution obtained using Eq.(\ref{eq:r2Sol}) and the numerical solution
of Eqs.(\ref{eq:T2})-(\ref{eq:T3}). In the left panel we show, in
the cases (a) and (b), that if $w_{\textrm{de}}$ enters the tracking
regime early, it evolves along the fixed points A, C, D and then stands
on the dS attractor. In this case, $w_{\textrm{de}}$ follows the
sequence $w_{\textrm{de}}=1/6-\alpha/6$ and $w_{\textrm{de}}=-7/3+\alpha/3$
in the radiation dominated era, then continues with $w_{de}=-2+\alpha/3$
in the matter dominated era, before it ends in the dS era with $w_{\textrm{de}}=-1+\alpha/3.$
On the other hand, if the tracking regime is reached later which is
shown in the cases (c) and (d), the sequence followed by $w_{\textrm{de}}$
is then given by the fixed points A, B, D before it stands on the
dS attractor. In this case, $w_{\textrm{de}}$ follows the sequence
$w_{\textrm{de}}=1/6-\alpha/6$ and $w_{\textrm{de}}=1/4-\alpha/6$
in the radiation and matter dominated small regime, before it reaches
$w_{\textrm{de}}=-1+\alpha/3$ in the dS era. We also note that, depending
on the initial conditions, $w_{\textrm{de}}$ always crosses earlier
or later the phantom divide line. We also show, that the tracker DE
EoS never reaches the small regime and follows the eras described by
the fixed points C, D and E and exhibits the following sequence: $w_{\textrm{de}}=-7/3+\alpha/3$
in radiation era, $w_{\textrm{de}}=-2+\alpha/3$ in matter era, and
$w_{\textrm{de}}=-1+\alpha/3$ in dS era, respectively. The right
panel of Fig.\ref{EoS_parameters} shows the evolution of $w_{\textrm{eff}}$.
As it can be seen, $w_{\textrm{\text{e}ff}}$ indifferently follows
one of the paths, A, C, D, E or A, B, D, E. It should be noted that
the behavior of $w_{de}$ in the matter era governed by the fixed
point D (anytime the tracking curve is reached) is not a source of
problems since $w_{eff}$ that governs the physical evolution of the
cosmological parameters realizes the standard cosmological eras of
the $\Lambda\textrm{CDM model, i.e, radiation era \ensuremath{\left(\Omega_{\textrm{r}}=1,\:w_{\textrm{eff}}=1/3\right),} }$matter
era $\left(\Omega_{\textrm{m}}=1,\:w_{eff}=0\right)$ and the dS era
$\left(\Omega_{\textrm{de}}=1,\:w_{\textrm{eff}}=-1\right).$ 

\begin{figure}
\centering \includegraphics[width=7cm,height=5cm]{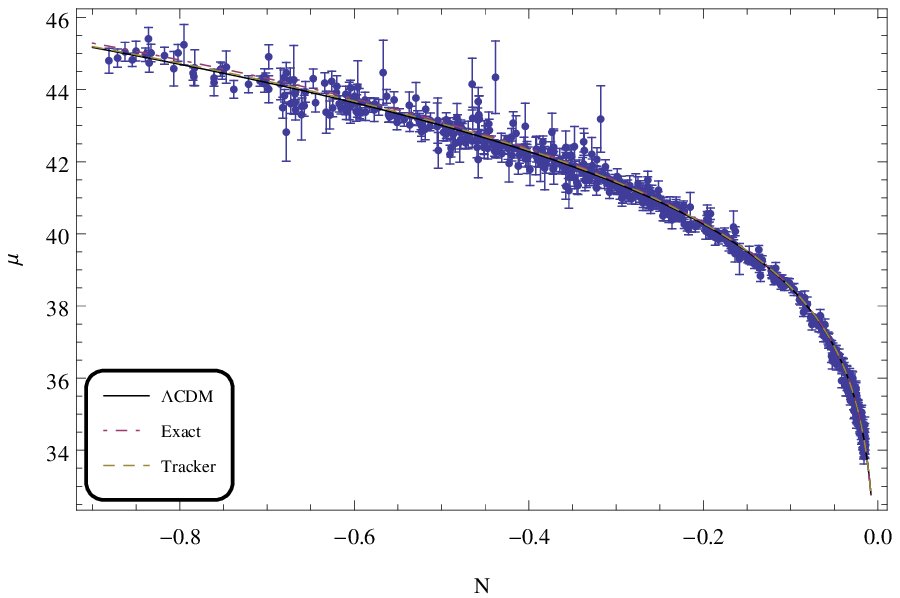}\hspace{0.5cm}
\hfill{}\includegraphics[width=7cm,height=5cm]{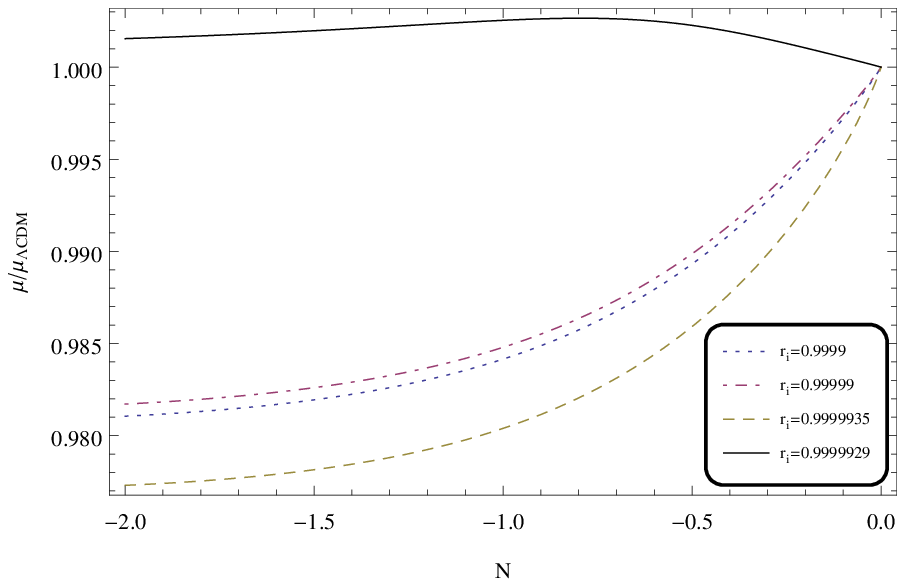}

\caption{\label{Lum_Dist_Diag} \textit{Left panel}: Distance modulus versus
$N$ for \textit{$\Omega_{ri}=0.9999929.$} \textit{Right panel}:
Comparison between the distance modulus in the interacting Galileon
model and $\Lambda$CDM for different initial conditions on the radiation
energy. In both panels we used $\alpha=-0.0075,\:r_{1i}=0.0715\times10^{-12},\:r_{2i}=7.17\times10^{-18},\:\Omega_{bi}=1.1013\times10^{-6}$.}
\end{figure}

\begin{itemize}
\item \textit{Distance luminosity and Hubble parameter:} 
\end{itemize}
In this paragraph we compare the background expansion history of the
interacting Galileon model with the $\Lambda$CDM using the distance
modulus and the Hubble parameter. We use the Union2.1 compilation
data for the Supernovae Ia type including 557 points \cite{union2.1}
and the latest cosmic chronometer $H(z$) data used in \cite{Basilakos-Nesseris}
and based on the datasets in \cite{Moresco et al,Zhang}. The distance
modulus $\mu\left(z\right)$, an observable quantity, is given in
terms of the luminosity distance $d_{\textrm{L}}=\left(1+z\right)\int_{0}^{z}\frac{du}{H\left(u\right)}$
by 
\begin{equation}
\mu\left(z\right)=5\log\left(\frac{d_{\textrm{L}}}{H_{0}\,\textrm{Mpc}}\right)+25
\end{equation}
where $H$ is given by 
\begin{equation}
H=H_{0}\left(\frac{r_{1}\left(0\right)}{r_{2}\left(N\right)}\right)^{5/4}\left(\frac{r_{2}\left(0\right)}{r_{2}\left(N\right)}\right)^{1/4}.\label{eq:Hub_param}
\end{equation}
In the right panel of Fig. \ref{Lum_Dist_Diag} we show examples of
the evolution of the distance modulus for fixed initial conditions
and coupling constant. The curves represent the scenarios including
the exact solution (dot-dashed), the approximate tracking solution
(dashed) and $\Lambda$CDM (solid), respectively. In the right panel
we fix the coupling constant to $\alpha=-0.0075$ and vary the initial
value of the radiation energy density while we keep the initial conditions
on $r_{1}$ and $r_{2}$ fixed and plot the ratio between the distance
modulus in our model and the distance modulus computed in the $\Lambda\textrm{CDM}$
model. As we can see the best result is reached for $\Omega_{r,i}=0.9999929.$
Exactly like the plots for the distance modulus, we see that the best
ratio between the Hubble parameter in our model and the Hubble parameter
in the $\Lambda\textrm{CDM}$ model with the same coupling constant
and initial conditions is also reached for $\Omega_{r,i}=0.9999929.$

\hfill{}

\begin{figure}
\centering \includegraphics[width=7cm,height=5cm]{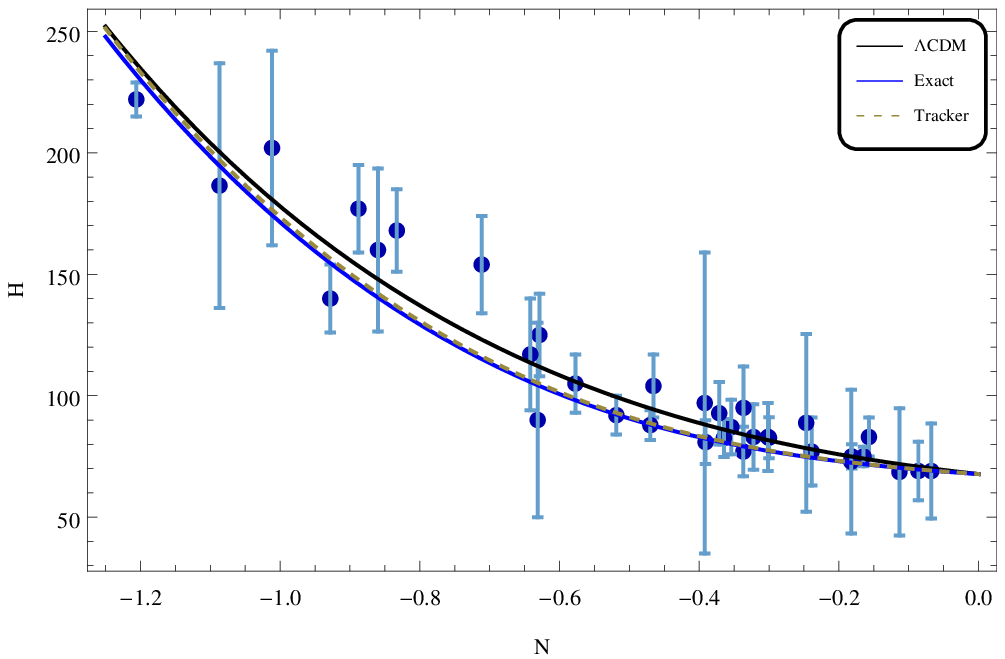}\hspace{0.5cm}
\hfill{}\includegraphics[width=7cm,height=5cm]{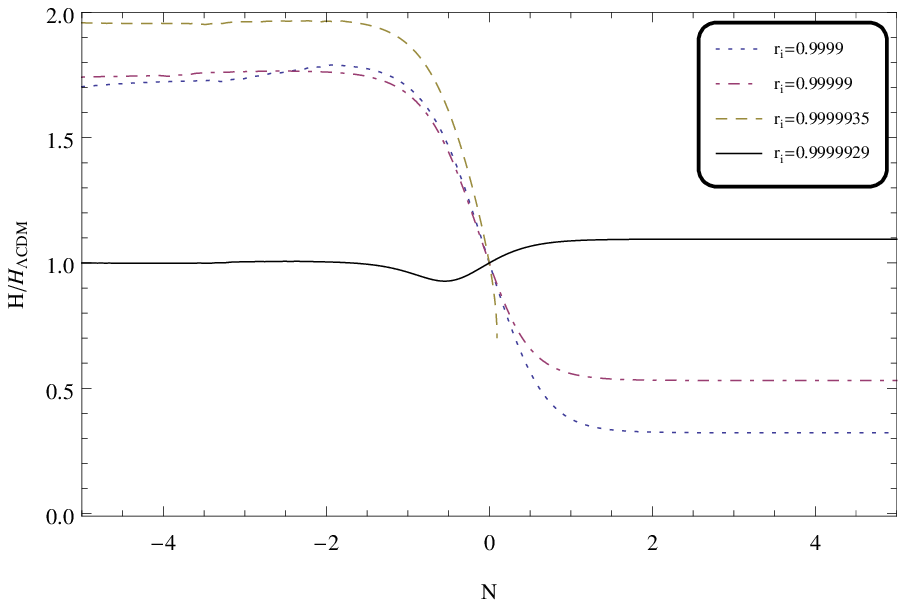}
\caption{\label{Hubble_Diag} \textit{Left panel:} Hubble parameter versus
$N$ for \textit{$\Omega_{ri}=0.9999929.$ Right panel}: Comparison
between the Hubble parameters in the interacting Galileon model and
$\Lambda$CDM. We use the same initial input values as Fig. \ref{Lum_Dist_Diag}.}
\end{figure}

\begin{figure}[tp]
\centering \includegraphics[width=7cm,height=5cm]{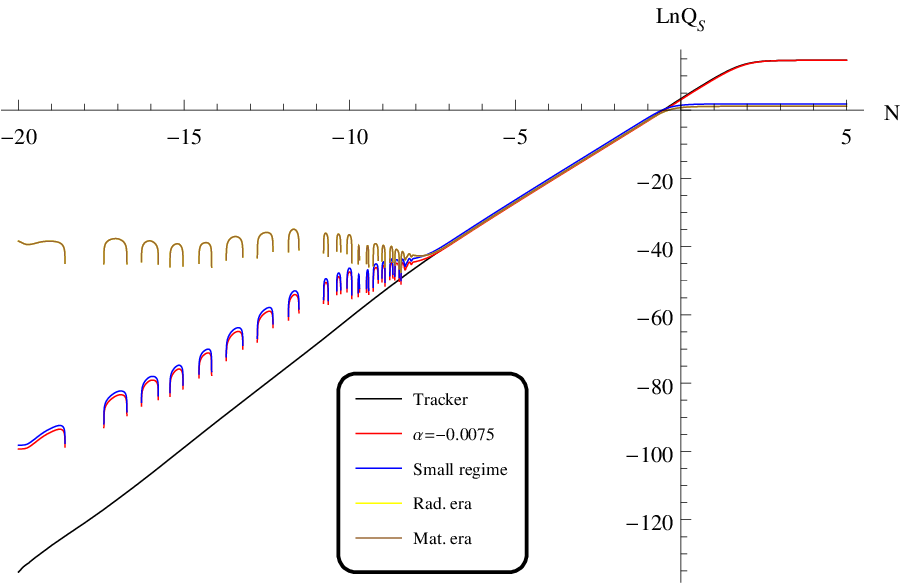} \hfill{}\includegraphics[width=7cm,height=5cm]{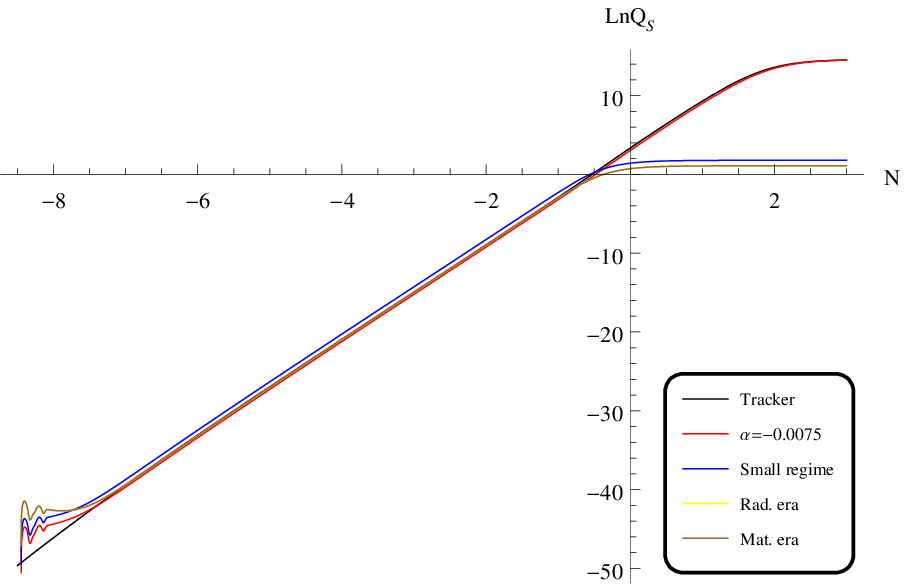}

\caption{\label{Qs-plot} \textit{Left panel}: Evolution of $Q_{S}$ versus
$N$ in different scenarios for $\alpha=-0.0075$, $\Omega_{ri}=0.9999929,\:\Omega_{bi}=1.10125\times10^{-6}$.
For the exact calculation the initial conditions are $r_{1i}=0.0715\times10^{-12},\:r_{2i}=7.17\times10^{-18}$,
while for the tracker we used \textit{$r_{2i}=0.71\times10^{-59}.$
Right panel}: zoom on the region $-8.5\leq N\leq2.5$.}
\end{figure}

\begin{figure}
\centering \includegraphics[width=7cm,height=5cm]{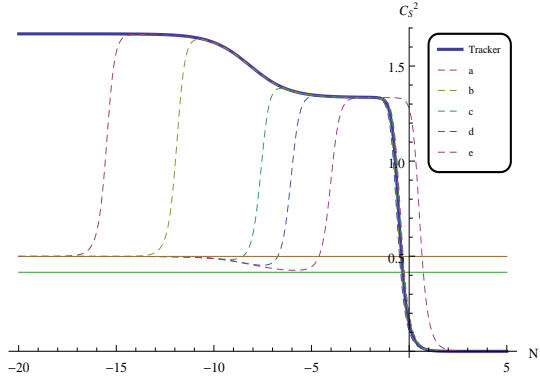}

\caption{\label{Cs-plot} Evolution of $C_{\textrm{S}}^{\text{2}}$ versus
$N$ for different scenarios for $\alpha=-0.0075$. The thin horizontal
lines are for $C_{S}^{\text{2}}\approx0.4987$ and $C_{S}^{2}\approx0.4154,$
in radiation and matter dominance eras, respectively. We use the same
initial input values as Fig. \ref{EoS_parameters} plus the curve
(e) with $r_{1i}=24.01\times10^{-17},\:r_{2i}=7.6\times10^{-1}$.}
\end{figure}

\begin{itemize}
\item \textit{No-ghost and Laplacian instabilities:} 
\end{itemize}
In Fig. \ref{Qs-plot} we plot the evolution of $Q_{\textrm{S}}$
in different scenarios: the exact numerical solution obtained by integrating
the autonomous system (\ref{eq:ode1}-\ref{eq:ode4}), the small regime
solution given by Eq.(\ref{eq:small_Qs}), the radiation and the matter
domination epochs obtained for $r_{2}$ small in the vicinity of the
fixed points C and D given by Eqs.(\ref{eq:rad_Qs}) and (\ref{eq:mat_Qs}),
respectively, and lastly the approximated tracking solution given
in Eq.(\ref{eq:Track_Qs}). As can be seen, the exact curve follows
exactly the small regime solution in the radiation era (\ref{eq:small_Qs}),
then merges at the onset of the matter domination era with the tracking
solution (\ref{eq:Track_Qs}), and the fixed points C and D solutions
given by (\ref{eq:rad_Qs}) and (\ref{eq:mat_Qs}), respectively,
before continuing alone along the tracking solution in the dS era.
The evolution of the propagation speed squared of the scalar modes
is plotted in Fig. \ref{Cs-plot}. In the small regime, $r_{1}\ll1$
and $r_{2}\ll1$, the exact curve follows the estimated values of
the scalar propagation speed given by (\ref{eq:small_speed_1}) and
(\ref{eq:small_speed_2}) in the radiation and matter small regime
eras, respectively. For $\alpha=-0.0075$, these values are given
by $C_{\textrm{S}}^{2}\approx0.4987$ and $C_{\textrm{S}}^{2}\approx0.4154$
for $\Omega_{r}=1$ and $\Omega_{r}=0,$ respectively. Then, the exact
curve follows the estimated values given in the eras around the points
C and D given by $C_{\textrm{S}}^{2}\approx1.667$ and $C_{\textrm{S}}^{2}\approx1.333$
for $\Omega_{r}=1$ and $\Omega_{r}=0,$ respectively, and finally
it follows the tracking solution at the onset of the matter era until
the dS era. However, this behavior is highly dependent on the initial
conditions. For example, in the cases (a) and (b) the exact curve
enters early the tracking regime and follows the points A, C, D and
ends in the dS era, while in the cases (c) and (d), the curve follows
later the tracking regime and follows the points A, B, D before it
ends in the dS era. This is exactly the behavior observed with $w_{\textrm{de}}.$
Finally, we can find a set of initial conditions, as shown in (e),
where the scalar mode starts sub-luminal, follows the approximated
tracker solution and becomes super-luminal temporarily during the eras
around the fixed points C and D, before it becomes sub-luminal in
the recent past until today where $C_{\textrm{S}}^{2}\left(z=0\right)\approx0.183$.
But, when it goes deeper in the dS era weak Laplacian instabilities
appear since $C_{\textrm{S}}^{2}\left(z\rightarrow-1\right)\approx-0.00041$.
As already noted in \cite{A. De Felice_PRL}, the fact that there
is a cosmological epoch where the propagation speed squared of scalar
modes is greater than one does not render the model unviable because
of the possibility for the absence of closed causal curves \cite{R. Gannoudji and M. Sami}.
Let us finally note that even we introduced an interaction between
the dark sectors the GWs propagate at the speed of light, and thus
the bounds imposed simultaneously by GW170817 and GRB170807A \cite{B. P. Abbott et al,B. P. Abbott et al. 2}
are trivialy satisfied .

\begin{figure}
\centering \includegraphics[width=5cm,height=5cm]{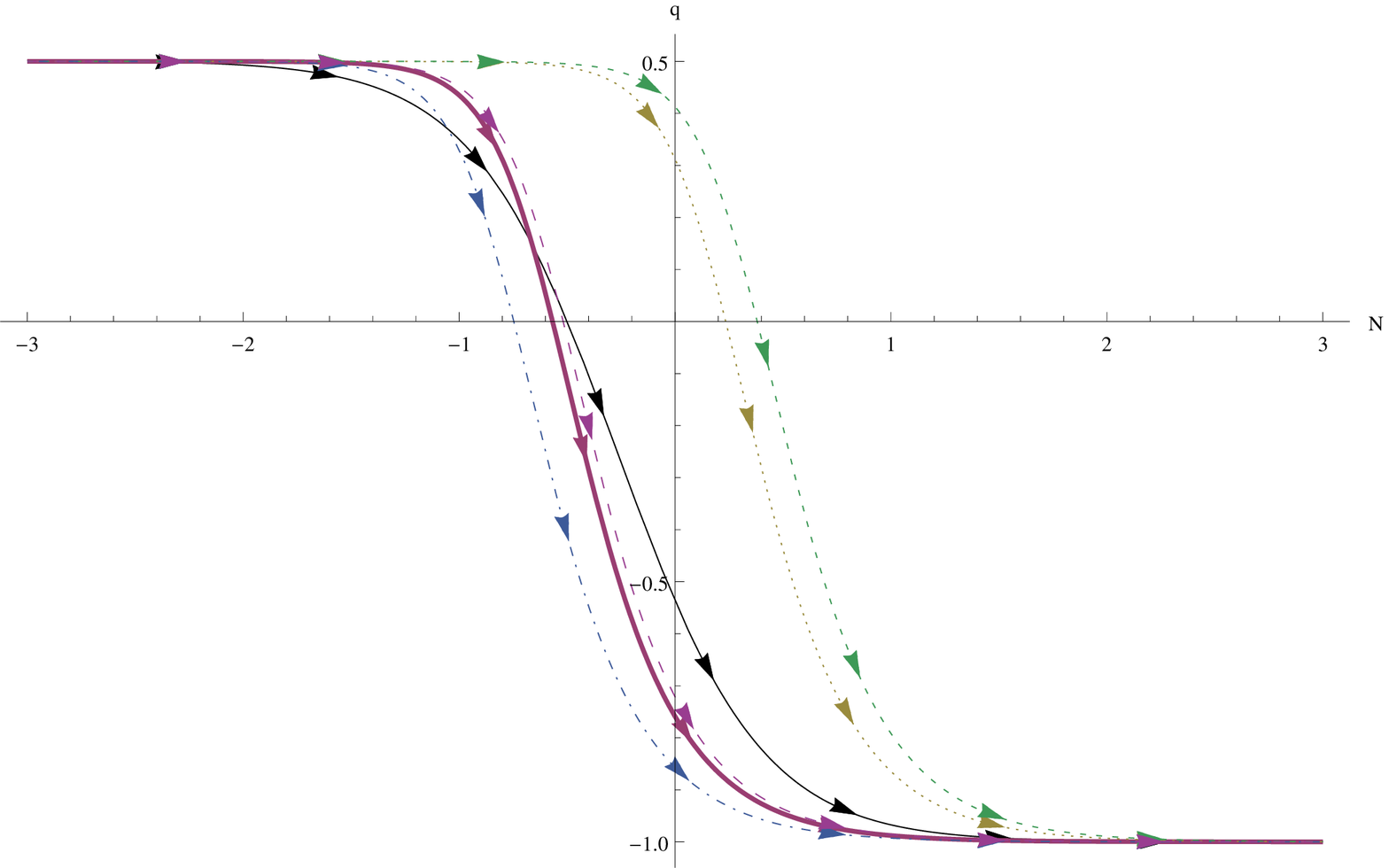} \hfill{}\includegraphics[width=5cm,height=5cm]{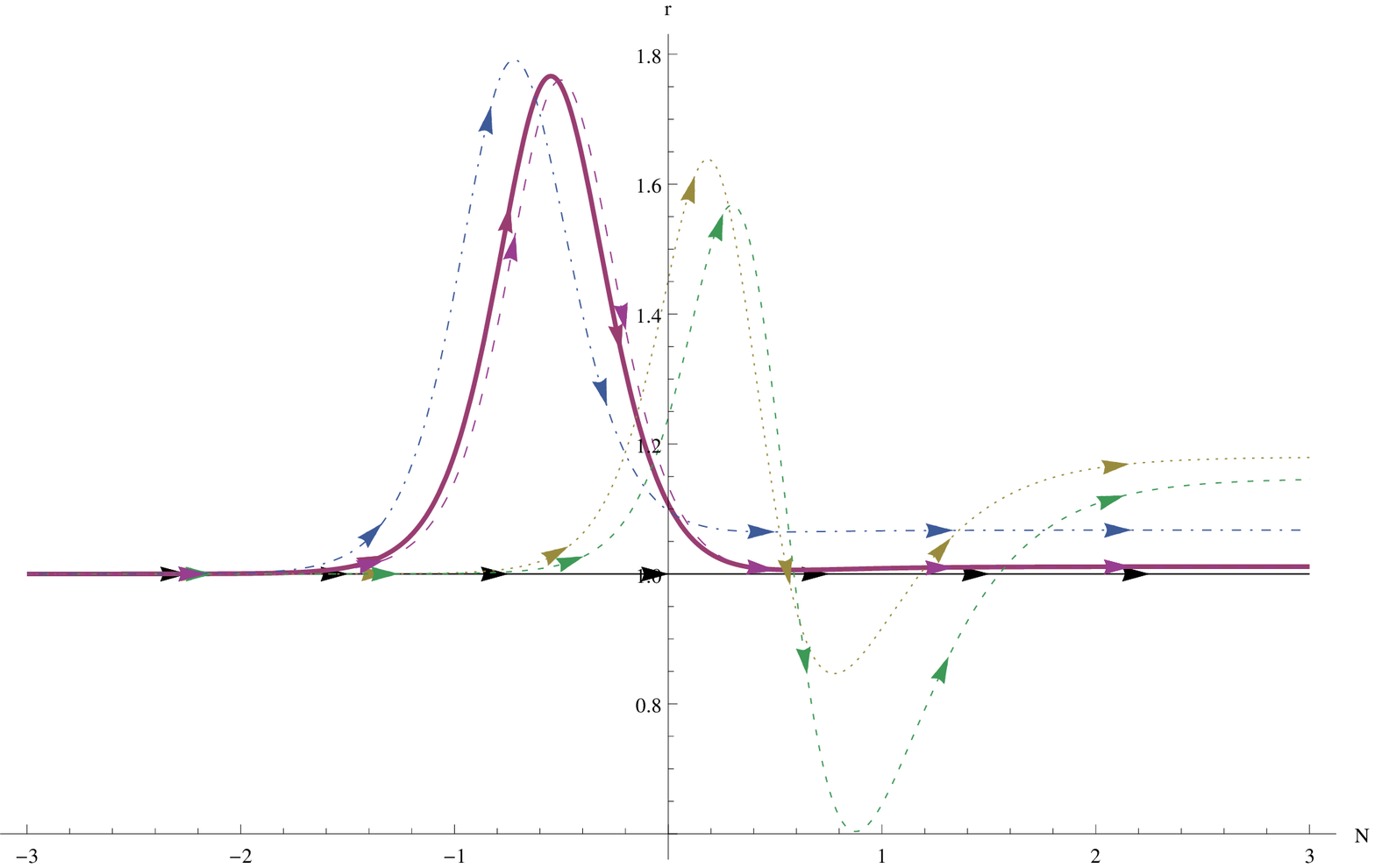}
\hfill{}\includegraphics[width=5cm,height=5cm]{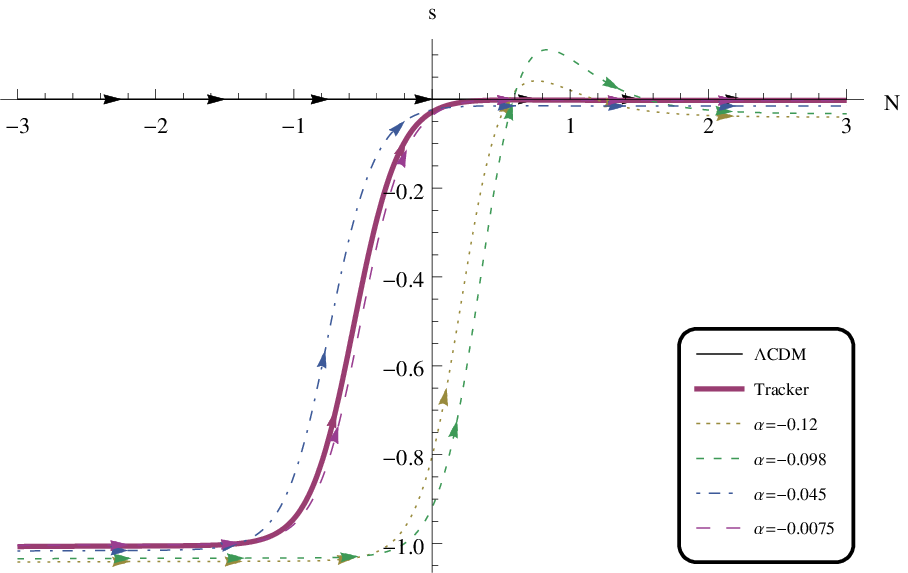}

\caption{\label{diag-1} Evolution of the deceleration parameter $q\left(N\right),$
and the statefinder parameters $\:r\left(N\right),\:s\left(N\right)$
for different values of the coupling $\alpha$ and the with the initial
conditions $r_{1i}=0.0715\times10^{-12},\:r_{2i}=7.17\times10^{-18},\:\Omega_{ri}=0.9999929,\:\Omega_{bi}=1.1013\times10^{-6}.$}
\end{figure}

\begin{figure}
\centering \includegraphics[width=5cm,height=5cm]{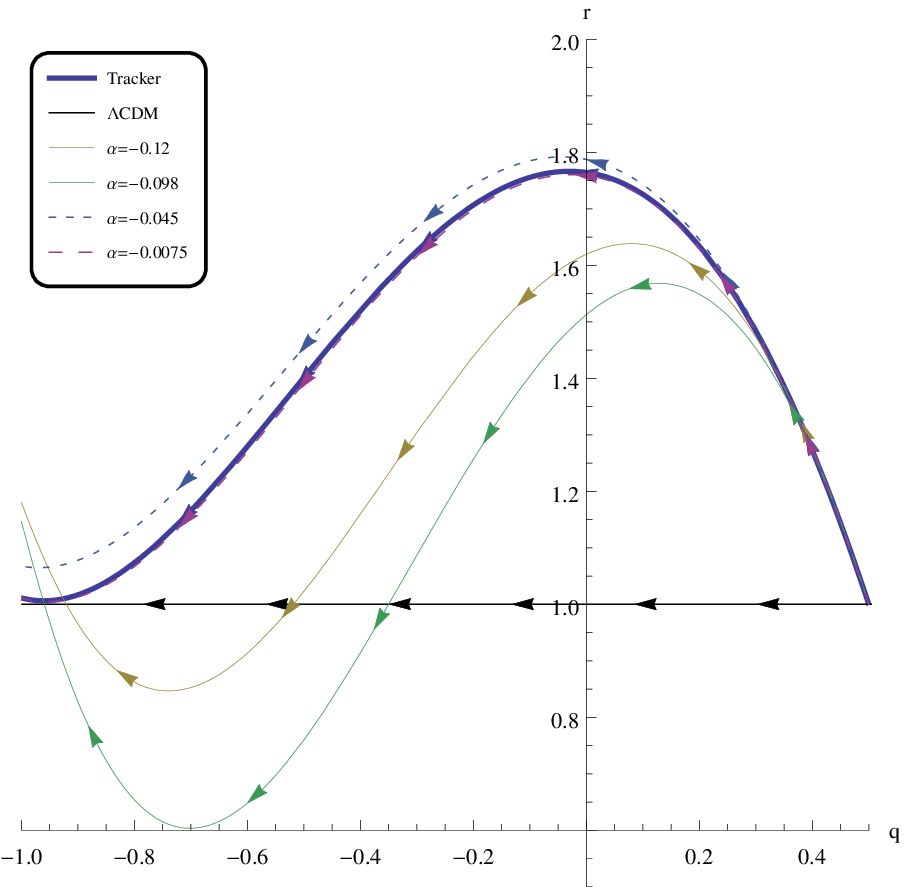} \hfill{}\includegraphics[width=5cm,height=5cm]{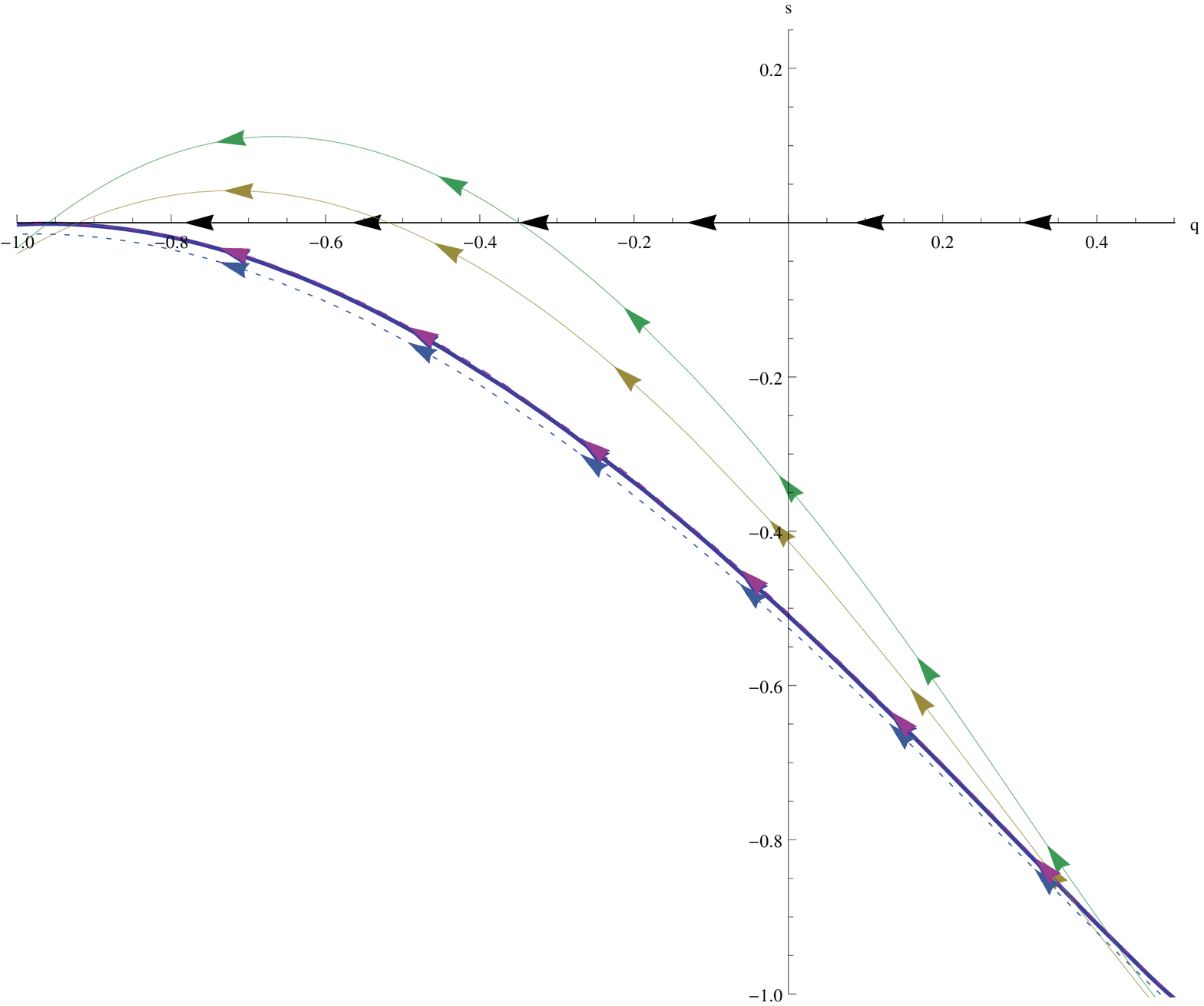}
\hfill{}\includegraphics[width=5cm,height=5cm]{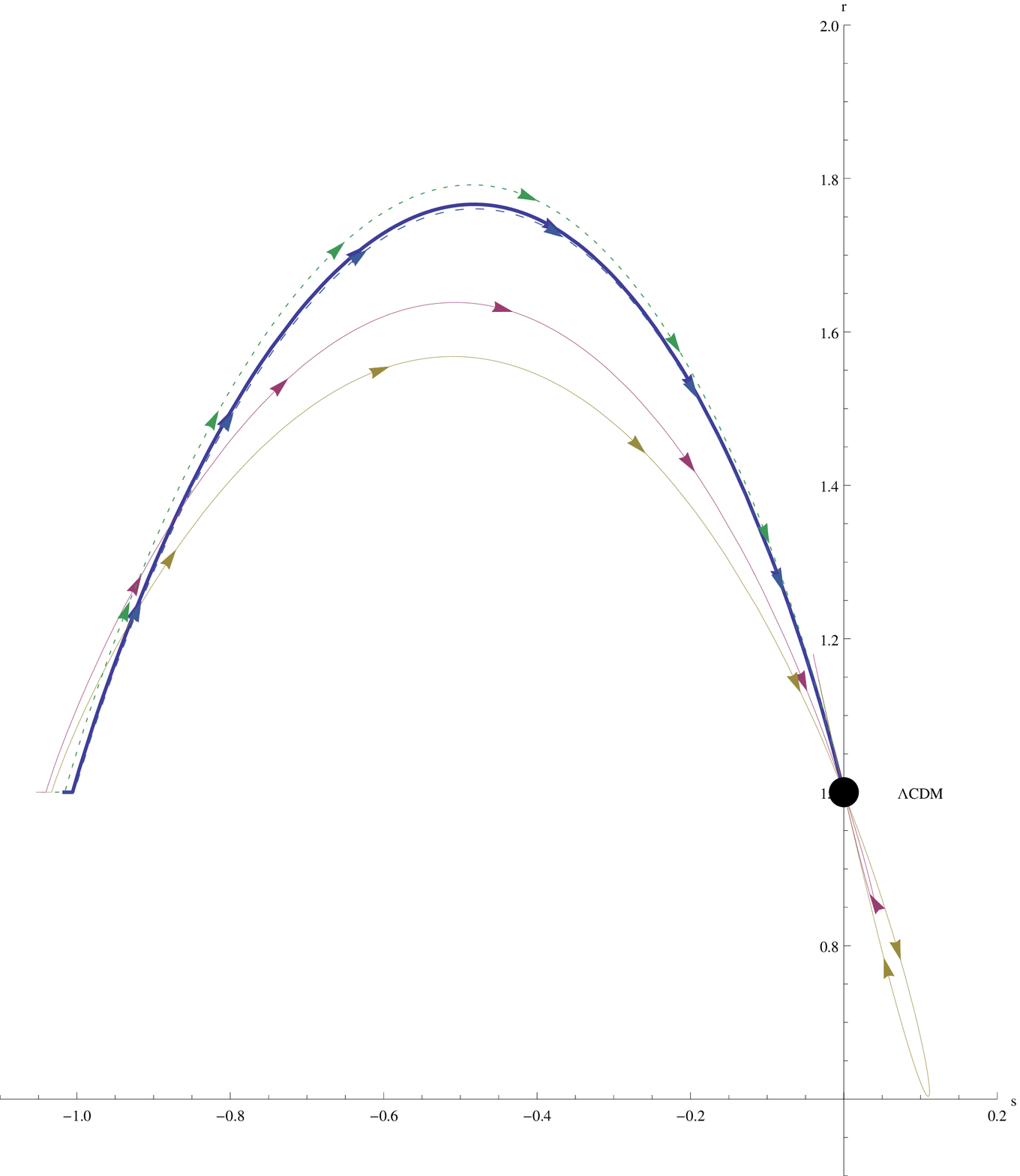}

\caption{\label{diag-2} Statefinder diagnostic in the $r-q,\:s-q$ and $r-s$
planes for the same parameters as Fig.\ref{diag-1}. }
\end{figure}

\begin{itemize}
\item \textit{Statefinder and $Om$ diagnostics}\label{sec:Statefinder-and-} 
\end{itemize}
In this section, we analyze the interacting Galileon model by the
statefinder diagnostic. In the $\Lambda\textrm{CDM}$ model, two fundamental
geometrical probes characterize the dynamics of the universe expansion,
the Hubble parameter $H$ and the deceleration parameter $q$ involving
only the scale factor, its first and second derivatives with respect
to time. However, with the great amount and increasing precision of
the cosmological observations data and the advent of a plethora of
dark energy models, the present question is how to discriminate between
these models, and how to quantify the distance to the $\Lambda\textrm{CDM}$
model. Since the Hubble and deceleration parameters are no longer
sensitive enough to discriminate between the different dark energy
models, Sahni et al.\cite{sahni} and Alam et al.\cite{Ulam} proposed
the so-called statefinder diagnostic based on higher derivatives of
the scale factor with respect to time and introduced new pair of parameters
$\left\{ r,s\right\} $ defined by 
\[
r=\frac{\dddot{a}}{aH^{3}}\quad s=\frac{r-1}{3\left(q-1/2\right)}
\]
where $q=-\ddot{a}/\left(aH^{2}\right)$ is the deceleration parameter.
It is expected from the design of future experiment \cite{J. Albert_1,J. Albert_2,Linder}
to obtain an estimate of these parameters. The $\Lambda\textrm{CDM}$
model corresponds to the fixed point $\left\{ 1,0\right\} $ in the
plane $\left\{ r,s\right\} .$ This feature, allows us to appreciate
the behavior of models of dark energy by measuring the distance between
them and the $\Lambda\textrm{CDM}$ fixed point. The full numerical
analysis of the statefinder parameters is shown in Figs.(\ref{diag-1})-(\ref{diag-2}).
Fig.\ref{diag-1} shows the time evolution of $q,$ $r$ and $s$
for different DE-DM coupling constant. In all panels the exact numerical
statefinder parameters (dashed curves) are compared to the tracking
solution (thick curve) and the $\Lambda\textrm{CDM}$ model results
(black thin curve). Apart deviations in the recent past due to DE-DM
interaction, we observe that for $\alpha=-0.0075$ the exact and the
tracker curves for $q$ and $r$ are practically indistinguishable
and are very close to the $\Lambda\textrm{CDM}$ curve. For the $s$
parameter and under the same setup the exact and the tracker curves
merge with the $\Lambda\textrm{CDM}$ curve only in the near future
and stand on in the dS era. In Fig. \ref{diag-2} and from left to
right, we illustrate the evolution of the pairs $\left\{ q,r\right\} ,\:\left\{ q,s\right\} $
and $\left\{ r,s\right\} .$ For the pairs $\left\{ q,r\right\} $
and $\left\{ q,s\right\} $ the exact and tracker curves for $\alpha=-0.0075$
which are indistinguishable start from the standard cold dark matter
(SCDM) point where $w_{eff}=0$ and $\left\{ q=1/2,\:r=1\right\} $
and converge toward the dS universe where $\left\{ q=-1,\:r=1\right\} .$
In the right panel all the curves start at the dS fixed point $\left\{ q=-1,\:r=1\right\} ,$
pass through the $\Lambda\textrm{CDM}$ fixed point $\left\{ r=1,\:s=0\right\} $,
and after some detours, the trajectories converge toward this point.
However, for $\alpha=-0.0075$, the exact and tracking curves converge
to the $\Lambda\textrm{CDM}$ fixed point without tracing visible
loops. The apparition of loops is suggested from the $r$-curves,
in the middle panel of Fig.\ref{diag-1}, which cross the $\Lambda\textrm{CDM}$
line from above in the recent past to below in the future (the curves
with $\alpha=-0.12$ and $\alpha=-0.098,$ respectively). As can
be seen, the $r$-curves which do not form loops never cross the $\Lambda\textrm{CDM}$
line. Similar conclusions can be traced from the $s$-curves in the
left panel of Fig.\ref{diag-1}.

\begin{figure}[H]
\centering \includegraphics[width=7cm,height=5cm]{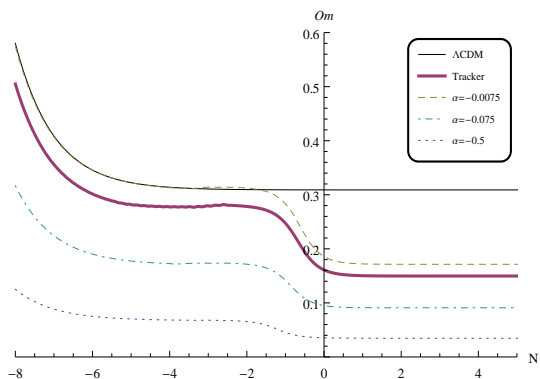}

\caption{\label{Om_diag} The $Om$ curves for different values of the coupling
constant. The initial conditions are $r_{1i}=0.0715\times10^{-12},\:r_{2i}=7.17\times10^{-18}.$ }
\end{figure}

The $Om$ diagnostic is an other very useful tool to discriminate
between different cosmological models. It involves only first order
derivative of the scale factor and is based on the following formula
\begin{equation}
Om\left(N\right)=\frac{E^{2}\left(N\right)-1}{e^{-3N}-1}
\end{equation}
where $E\left(N\right)=H\left(N\right)/H_{0}$ is given by (\ref{eq:Hub_param}).
We have included the radiation contribution in our analysis. For non-interacting
dark energy models with constant $w_{de},$ a positive slope of $Om$
corresponds to phantom like behavior $\left(w_{de}<-1\right)$, a
negative slope corresponds to quintessence like behavior $\left(w_{de}>-1\right),$
and zero slope to the cosmological constant. In Fig.(\ref{Om_diag})
we show the $Om$ curves for the exact, tracker and $\Lambda$CDM
calculations. For $\alpha=-0.0075$, the exact curve starts following
the $\Lambda$CDM curve at the onset of the matter domination epoch,
then diverge from the $\Lambda$CDM curve in the recent past until
it becomes constant in the dS era. The difference in the trajectory
evolution between the $\Lambda$CDM and the interacting Galileon model
at the onset of DE domination epoch can be explained by the DE-DM
interaction which only becomes relevant in the recent past. As we
can see we have practically always a positive curvature signaling a
phantom behavior. 

\section{Conclusion\label{sec:Conclusion}}

In our present work, we considered the DE-DM interaction in the cubic
Galileon model where the interaction term is proportional to Hubble
parameter and Galileon dark energy in the form $Q=\alpha H\rho_{de}.$
We first assumed the existence of a dS cosmological era, which allowed
us to write the free parameters in the Galileon Lagrangian in terms
of the coupling constant and thus we reduced the dimension of the
parameter space. Using appropriate dimensionless variables, $r_{1},\:r_{2}$
besides the radiation and baryonic energy densities, the field equations
in flat FLRW background have been converted into an autonomous system
of first order differential equations. The dynamical system analysis
of the model has revealed the usual physical phase space reproducing
the different cosmological eras, radiation domination era, matter
domination era and the dark energy domination era, and possess a de
Sitter expansion fate in the future.

We carried out the exact numerical integration of the dynamical systems
of equations (\ref{eq:de1}-\ref{eq:de4}) and analyzed the evolution
of the background quantities like the energy densities $\Omega_{i}$,
the dark energy EoS $w_{de}$ and the total effective EoS parameters
$w_{eff}$, as well as $C_{S}^{2}$ and $Q_{S}$. The conditions on
the tensor modes being automatically satisfied, we derived the conditions
for the avoidance of ghosts and Laplacian instabilities associated
with scalar in terms of the dynamical variables and the coupling constant.
The stability analysis of the critical points and the conditions for
the avoidance of ghosts and Laplacian instabilities constrained the
coupling constant to negative values. The existence of a stable attractor
in the dS era allowed us to construct an approximated tracking solution
which mimics the exact solution, particularly in the recent past before
reaching the dS era. We have shown that the exact curves representing
the dark energy EoS, $w_{de},$ converge to the tracker solution at
early or later times depending on the initial conditions of the dynamical
variables. Particularly, we obtained that along the tracker solution,
$w_{de}$ follows paths belonging to the eras described by the fixed
points A, C, D and E and exhibits peculiar behavior: $w_{de}=-7/3+\alpha/3$
in the radiation era, $w_{de}=-2+\alpha/3$ in the matter era, and
$w_{de}=-1+\alpha/3$ in the dS era. However, depending on the initial
conditions on $r_{1}$ and $r_{2}$, the exact $w_{de}$ follows one
of the sequences: A, B, D, E at earlier times for $r_{1i}$ small,
or A, C, D, E at later time for $r_{2i}$ small . As we can see the
tracker solution is not compatible with observations since $w_{de}$
is far away from $-1$ in the matter era \cite{S. Nesseris and A. De Felice}.
The same problem was met in the quintic Galileon model
and solved in the framework of the extended Galileon model \cite{A. De Felice and S. Tsujikawa}.
Then, it will be of interest to investigate the DE-DM interaction
in the extended Galileon model and evaluate the combined effects of
$\alpha$ and $s$ on the viability of the tracking solution. 

Concerning the conditions for the avoidance of ghosts and Laplacian
instabilities associated with scalar perturbations, we found that
the exact curve follows exactly the small regime curve in the radiation
dominated era, then merges with the tracking and the solutions around
the fixed points C and D, and finally follows the tracking solution
alone at in the dS era. For the evolution of the propagation speed
squared of the scalar mode, we found that the mode starts super-luminal
during the eras around the fixed points C and D, before it becomes
sub-luminal in the recent past until entering the dS point where it
takes the value $C_{\textrm{S}}^{2}\approx4.16\times10^{-5}$ for
$\alpha=-0.0075$. In all cosmological epochs, $Q_{S}$ remains positive,
while $C_{S}^{2}$ which is positive until today, $C_{\textrm{S}}^{2}\left(z=0\right)\approx0.183,$
and in the future with $C_{\textrm{S}}^{2}\left(z\approx-0.85\right)\approx7.35176\times10^{-6},$
becomes weakly negative in the depth of the dS epoch where $C_{\textrm{S}}^{2}\left(z\rightarrow-1\right)\approx-0.00041$.

We have also compared the theoretical distance modulus and Hubble
parameter with the observation data. We found best agreement for a
set of initial conditions and coupling constant, and that for fixed
initial conditions on $r_{1}$ and $r_{2}$ and $\alpha=-0.0075,$
the best agreement between the exact results of the interacting cubic
Galileon model and the $\Lambda$CDM model is reached for $\Omega_{ri}=0.9999929.$

Employing the statefinder diagnostic, we found that the interacting
cubic Galileon model can be distinguished from the $\Lambda$CDM model.
Particularly, the parameter $s$ shows a completely different behavior
starting in the matter domination era until the recent past compared
to the $\Lambda$CDM model. We have also shown that all the trajectories
start from the SCDM fixed point $\left\{ q=1/2,\:r=1\right\} $ but
do not converge all to the dS fixed point $\left\{ q=-1,\:r=1\right\} $
due to their phantom evolution. However for $\alpha=-0.0075$, the
evolution of exact and the tracking curves are indistinguishable and
the approach to the dS point becomes more precise. In the $r-s$ plane,
all the trajectories start approximatively at the point $\left\{ s=-1,\:r=1\right\} $,
pass through the $\Lambda$CDM fixed point, and depending on the
value of the coupling constant and the initial conditions, trace loops
before converging to the $\Lambda$CDM point. In addition we used
the $Om$ diagnostic, where we showed that the exact and tracking
curves are indistinguishable and begin to diverge from the $\Lambda$CDM
model at the onset of the late time acceleration era.

In summary, we have shown that the interacting Cubic Galileon model
constitute a viable dark energy model for a DE-DM interaction function
in the form $Q=\alpha H\rho_{de}$, since it is not plagued by a negative
evolution of dark matter density in the dS era, unlike the case of
constant $w_{\textrm{de}}$. However, it suffers from the appearance
of soft Laplacian instabilities in the far future of the dS era. A
full treatment of the evolution of matter density perturbations and
gravitational potentials and a joint data analysis using observational
data such Supernovae type Ia, BAO distance measurements, Weak gravitational
lensing, and CMB observations are necessary  to place constraints
on the coupling constant for the exact and the approximated tracker
solutions. We leave these issues for forcoming works.

\end{document}